\documentclass[11pt,a4paper]{article}
\usepackage{jheppub}
\usepackage[utf8]{inputenc}
\usepackage{graphicx,fancybox,float,comment}
\usepackage{units}
\usepackage{subcaption}
\usepackage{multirow,array,arydshln}
\usepackage{xcolor}
\usepackage{amsmath,amssymb}
\usepackage{cleveref}
\usepackage{yfonts}
\usepackage{tensor}
\usepackage{paralist}
\usepackage{braket}
\usepackage{tensor,mathrsfs}
\usepackage{dsfont}
\usepackage{ulem}
\DeclareMathAlphabet{\mathpzc}{OT1}{pzc}{m}{it}

\newcommand{\be}{\begin{eqnarray}}
\newcommand{\ee}{\end{eqnarray}}

\title{
Topological confinement in Skyrme holography
}

\author[a]{Casey Cartwright}
\emailAdd{cccartwright@crimson.ua.edu}

\author[a]{Benjamin Harms} 
\emailAdd{bharms@ua.edu}

\author[a]{Matthias Kaminski} 
\emailAdd{mski@ua.edu}

\author[b]{Ronny Thomale}
\emailAdd{rthomale@physik.uni-wuerzburg.de}

\affiliation[a]{Department of Physics and Astronomy, University of Alabama, 
\\ 514 University Boulevard,
Tuscaloosa, AL 35487, USA}

\affiliation[b]{Institut f{\"u}r Theoretische Festk{\"o}rperphysik (TP1) \\ Julius-Maximilians-Universit{\"a}t W{\"u}rzburg, Am Hubland, D-97074 W{\"u}rzburg, Germany}

\newcommand{\exd}{\mathrm{d}}

\definecolor{amaranth}{rgb}{0.9, 0.17, 0.31}

\newcommand{\vev}[1]{\braket{#1}}

\newcommand{\tr}{\text{Tr}}

\begin{document}

\abstract{
We study phase transitions in five-dimensional Einstein Gravity with a negative cosmological constant, coupled to a Skyrme matter field.
These transitions are topological generalizations of the Hawking-Page transition between thermal Anti de Sitter (AdS) spacetime and an AdS black hole. 
Phases are characterized by a topological number associated with the Skyrme field configuration. 
Depending on that topological number and on the Skyrme coupling strength, there occur transitions between those phases at two, one, or no value(s) of the temperature. 
Through the holographic (AdS/CFT) correspondence, these solutions are dual to topologically non-trivial states in a conformal field theory (CFT) with an SU(2)-symmetry, which support either confined or deconfined (quasi-)particles at strong coupling. 
We compare to similar known phase transitions, and discuss  potential applications to confinement in topological phases of condensed matter and the quark-gluon plasma. }

\maketitle

\section{Introduction} 
\label{sec:intro}
The study of phase transitions in the context of topological black holes in five-dimensional anti-de Sitter spacetime is important because of the duality between such black holes and the four-dimensional conformal field theories which live on the boundary. An understanding of the dynamics of phase transitions within gravitational systems in the bulk can potentially elucidate the dual transitions which occur in the strongly-coupled system on the boundary. In \cite{Cartwright:2020yoc} an analytic solution of the Einstein equations for such a black hole endowed with a Skyrme field was obtained and shown to be holographically~\cite{Maldacena:1997re} dual to an $\mathcal{N}\,=\,4$ Super-Yang-Mills (SYM) gauge field theory. For a large Skyrme coupling constant the Skyrme field can significantly alter the temperature at which the gravitational system undergoes a Hawking - Page (H - P) transition between thermal AdS and an AdS black hole.  As we show in section~\ref{sec:gravity}, the size of the coupling also determines whether the number of H - P transitions is zero, one, or two at a given topological charge value. 
Via the holographic (AdS/CFT) correspondence, the H - P transition was found to be dual to the (de-)confinement transition of the $\mathcal{N}=4$ SYM field theory~\cite{Witten:1998zw}. In~\cite{Cartwright:2020yoc}, the Skyrme coupling and Skyrme field were shown to correspond to $SU(2)$-gauge fields and their coupling to $\mathcal{N}=4$ SYM theory, respectively. 

The results of our work form the foundation of a more generalized treatment of topological phases of matter at strong coupling, as our technique allows for an analytic study of such transitions. The analytic solutions we have found can, via AdS/CFT duality, be applied  to other physical systems in which topological phase transitions occur.  
Our mapping of a topological black hole to its CFT analogue allows the analysis of (de-)confinement transitions which are decorated with a non-Abelian gauge field texture of potentially topologically non-trivial character. The intertwining of confinement and topology is ubiquitous in condensed matter. For quantum spin chains, the spinons as elementary quasiparticles can develop confinement to form a fully symmetric yet topologically non-trivial ground state for integer spins, or retain their deconfined character in the spin fluid phase of half-integer spin chains~\cite{haldane2016ground}. Likewise, certain spin liquids in 2+1 dimensional Mott insulators can be interpreted as the deconfined limit of an SU(2) gauge theory~\cite{PhysRevB.65.165113}, bearing similarities to certain parton mean fields describing fractional quantum Hall systems. Quantum phase transitions further allow for emergent gauge fields in their critical theory that cannot be reconciled by the adjacent order parameter~\cite{doi:10.1126/science.1091806}, but instead embody new elementary excitations with potentially topological features.

Another application could be (de-)confined topological phases and transitions between them, potentially occurring in heavy ion collisions. In that particle physics context, topological field configurations are believed to play a key role in the generation of axial charge asymmetry in heavy ion collisions, and potentially to the matter/anti-matter asymmetry of the universe~\cite{Kharzeev:2004ey,Ghosh:2021naw,Cartwright:2021maz}.

In this paper we carry out a thermodynamic analysis of the phase transitions which can occur in our topological black hole. We begin with a brief discussion in section~\ref{sec:gravity} where describe our topological black hole. We analyze the H - P transitions which can occur in our model and discuss the stability our gravitational system. The holographic interpretation of the results of section~\ref{sec:gravity} is given in section~\ref{sec:holo}.  In section~\ref{sec:transitions}, we compare and contrast the phase transitions described by our model to the phase transitions which occur in other physical systems.

\section{Topological black holes with Skyrmion hair}
\label{sec:gravity}
\subsection{Our model}
The Skyrmion action in five dimensions is given by\footnote{In our conventions, Greek indices $\mu,
\, \nu,\, ...\in \{0, 1, 2, 3,4 \}$ range over the five-dimensional gravity bulk coordinates, Latin indices $m,\, n, ...\in \{0, 1, 2, 3 \}$ range over the four-dimensional boundary field theory coordinates.}
\be \label{eq:skyrmeAction}
S = \int d^5x \sqrt{-g}\left(\frac{R-2 \Lambda}{16\pi G_5}+\frac{f_{\pi}}{16\pi}K_{\mu}K^{\mu}+\frac{1}{32\, \tilde{e}^2}\text{Tr}\left(\left[ K_{\mu}, K_{\nu}\right] \left[ K^{\mu}, K^{\nu}\right]\right),\right)  \label{eq:lagrange},
\ee
in which $R$ is the Ricci scalar, $\Lambda=-6/L^2$ is the cosmological constant in $AdS_5$, $L$ is the $AdS_5$ radius and $G_5$ is the five-dimensional gravitational constant, while $\tilde{e}$ is the Skyrme coupling constant. The Skyrme field $U$ is an $SU(2)$-valued Lorentz scalar and has been introduced into the action via the $SU(2)$-valued Lorentz four-vector ${K_{\mu}\,=\,U^{-1}\partial_{\mu} U}$. The Skyrme field tensor is defined as
\be 
F_{\mu \nu}\,=\,\left[ K_{\mu}, K_{\nu}\right]
\ee
In the following sections we will work with $\tilde{e}=8\pi G_5 e$ where $e$ is a dimensionless coupling, measuring the strength of the Skyrme coupling compared to the gravitational coupling. \\

\noindent \textbf{Topological solution: }In \cite{Cartwright:2020yoc} an analytic solution to the Einstein-Skyrme equations of motions was discovered. There, the authors demonstrated that the solution could be mapped to a solution of the massless $SU(2)$ Einstein-Yang-Mills theory. Furthermore it was demonstrated that the configuration of the equivalent $SU(2)$ gauge field is topologically non-trivial, characterized by a non-vanishing topological charge. This charge counts the number of times the mapping wraps the internal $SU(2)$ group manifold. Here, we briefly review this solution, which in general can be written as 
\begin{equation}
    U=e^{i n \chi v^i \tau^i},\label{eq:top_Ansatz}
\end{equation}
where $v^iv^i=1$ and $n\in \mathbb{Z}$ and $\tau_a$ are the Pauli matrices
\begin{equation}
\tau_1=\begin{pmatrix} 0& 1 \\ 1 & 0 \end{pmatrix},\quad \tau_2=\begin{pmatrix} 0& -i \\ i & 0 \end{pmatrix}, \quad \tau_3=\begin{pmatrix} 1 &0 \\ 0 &-1\end{pmatrix}, 
\end{equation}
such that $\text{tr}\{\tau_1 \tau_2 \tau_3\} = 2 i$ . The integer $n$ quantifies how many times the mapping wraps the internal $S^3$. The coordinates chosen here are $(t,\psi,\theta,\phi,r)$,
where $\psi\in (0,\pi)$, $\theta\in (0,\pi)$, $\phi\in (0,2\pi)$, $t$ is the temporal coordinate and $r$ is the bulk $AdS$ radial coordinate. A particularly useful form of the unit vector $v$ is given by
\begin{equation}
    v=(\cos (\theta ),\sin (\theta ) \cos (\phi ),\sin (\theta ) \sin (\phi )),\label{eq:unit}
\end{equation}
and we choose the Skyrmion field $\chi$ such that  $\chi=\psi$. For $n=1$ this results in a standard mapping of the unit three-sphere into the $SU(2)$ gauge manifold. The spacetime metric is taken to be
\begin{equation}
    ds^2=\frac{1}{A(r)}\exd r^2-A(r)\exd t^2+r^2\left(h_1(\psi)\exd \psi^2+h_2(\psi)(\exd\theta^2+\sin(\theta)^2\exd\phi^2)\right).
\end{equation}
The Einstein-Skyrme equations of motion at vanishing pion coupling, $f_\pi=0$, result in three independent equations for $h_1,h_2$ and $A$. The solutions of these equations are given by
  \begin{equation}
    h_1(\psi)=n^2,\quad h_2(\psi)=\sin(n\psi),\quad A(r)=\frac{1}{e^2 r^2}+\frac{\log (r)}{e^2 r^2}+\frac{r^2}{L^2}-\frac{m_t}{r^2}+1 \label{eq:Topo_non_trivial_Metric}\, ,
\end{equation} 
where $m_t$ may be thought of as the black hole mass.
As stated in~\cite{Cartwright:2020qov}, using the solutions in eq.~(\ref{eq:Topo_non_trivial_Metric}) along with $K_{\mu}=U\partial_\mu U$ for the $U$ given in eq.~(\ref{eq:top_Ansatz}) and eq.~(\ref{eq:unit}) the Skyrme equations of motion, along with the Yang-Mills equations of motion can be shown to be trivially satisfied. \\

\begin{figure}[H]
    \centering
   \includegraphics[width=0.49\textwidth]{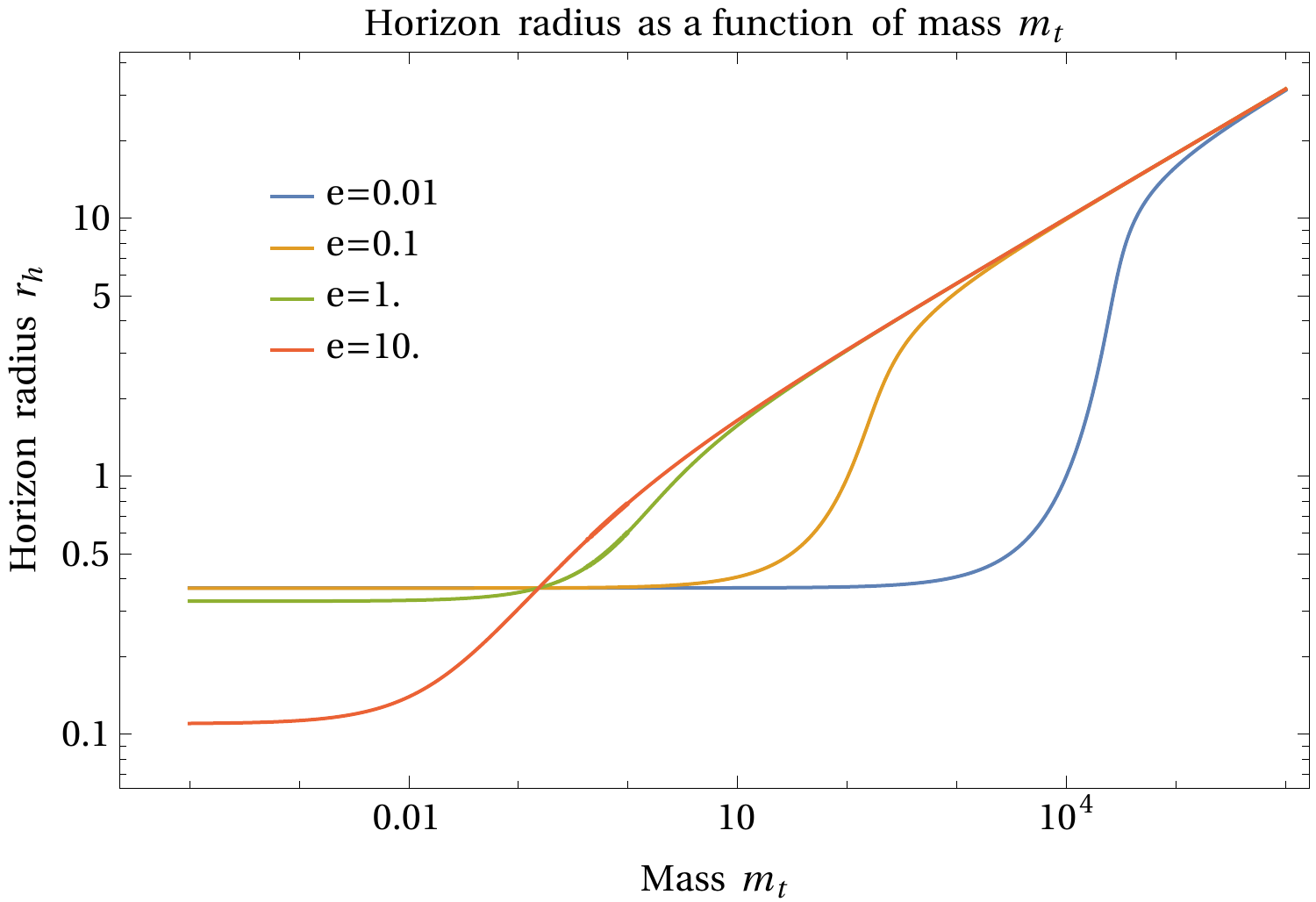}\hfill \includegraphics[width=0.49\textwidth]{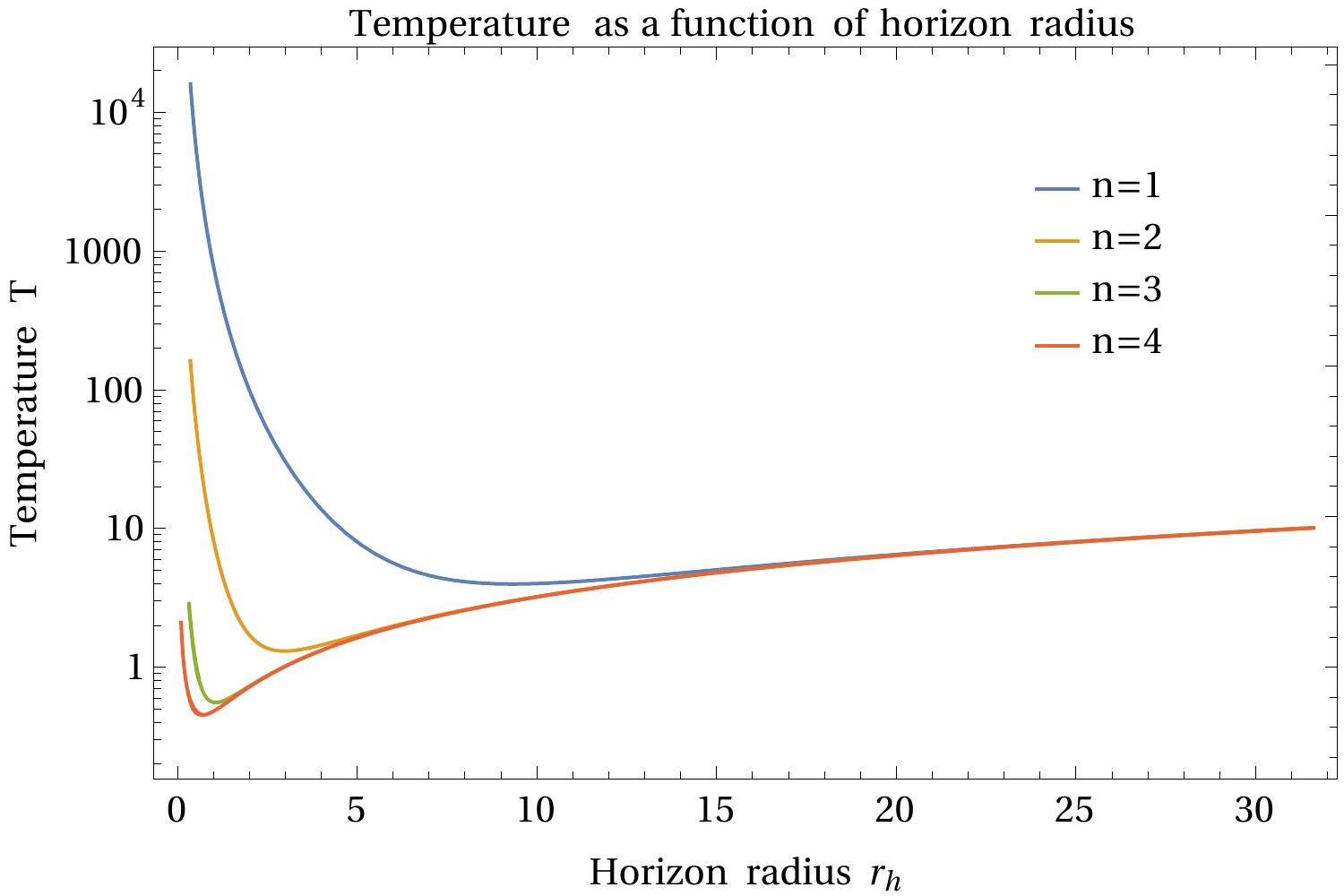}
    \caption{\textit{Left:}  The horizon radius of the solution as a function of the black hole mass $m_t$. \textit{Right:} The temperature $T$ of the solution as a function of the horizon radius. 
    \label{fig:temperature_v_horizon_radius}}
\end{figure}

\noindent\textbf{Temperature:} The solution has a horizon at the location $A(r_h)=0$. That equation is transcendental and can only be solved for $r_h$ numerically. 
The horizon radius of solutions (at various values of $e$) is displayed in the left panel of figure~\ref{fig:temperature_v_horizon_radius} as a function of the mass $m_t$. 
The temperature of such solutions is given by the standard formula
\begin{equation}
    T=\frac{1}{4\pi}|A'(r_h)| \, ,
\end{equation}
where the temperature as a function of the horizon radius is displayed in the right panel of figure~\ref{fig:temperature_v_horizon_radius}. Notably, there exists a minimal temperature $T_{\text{min}}$ for solutions of each winding number $n$ and Skyrme coupling $e$, indicating that only solutions with temperatures $T\ge T_{\text{min}}$ exist.  

The limit 
$e\rightarrow 0$ corresponds to an infinitely strong effect of the Skyrmion as a matter source in the bulk geometry. 
We note that the black hole of this configuration has a finite horizon radius determined by
\begin{equation}
A(r_h)=0=\frac{1}{e^2 r_h^2}+\frac{\log (r_h)}{e^2 r_h^2}+\frac{r_h^2}{L^2}-\frac{m_t}{r_h^2}+1 \hookrightarrow \frac{1}{ r_h^2}+\frac{\log (r_h)}{r_h^2}+\frac{e^2r_h^2}{L^2}-\frac{e^2\,m_t}{r_h^2}+e^2=0 \, ,
\end{equation}
Taking the limit $e\rightarrow 0$, we find 
\begin{equation}
    \frac{1}{ r_h^2}+\frac{\log (r_h)}{r_h^2}=0 \rightarrow r_h=\text{exp({-1})}\approx 0.368\, .
\end{equation}
The black hole attains a finite horizon radius as $e\rightarrow 0$, and as this occurs, $T\rightarrow \infty$. \\\\

\begin{figure}[htb]
\begin{center}
    \includegraphics[width=0.49\textwidth]{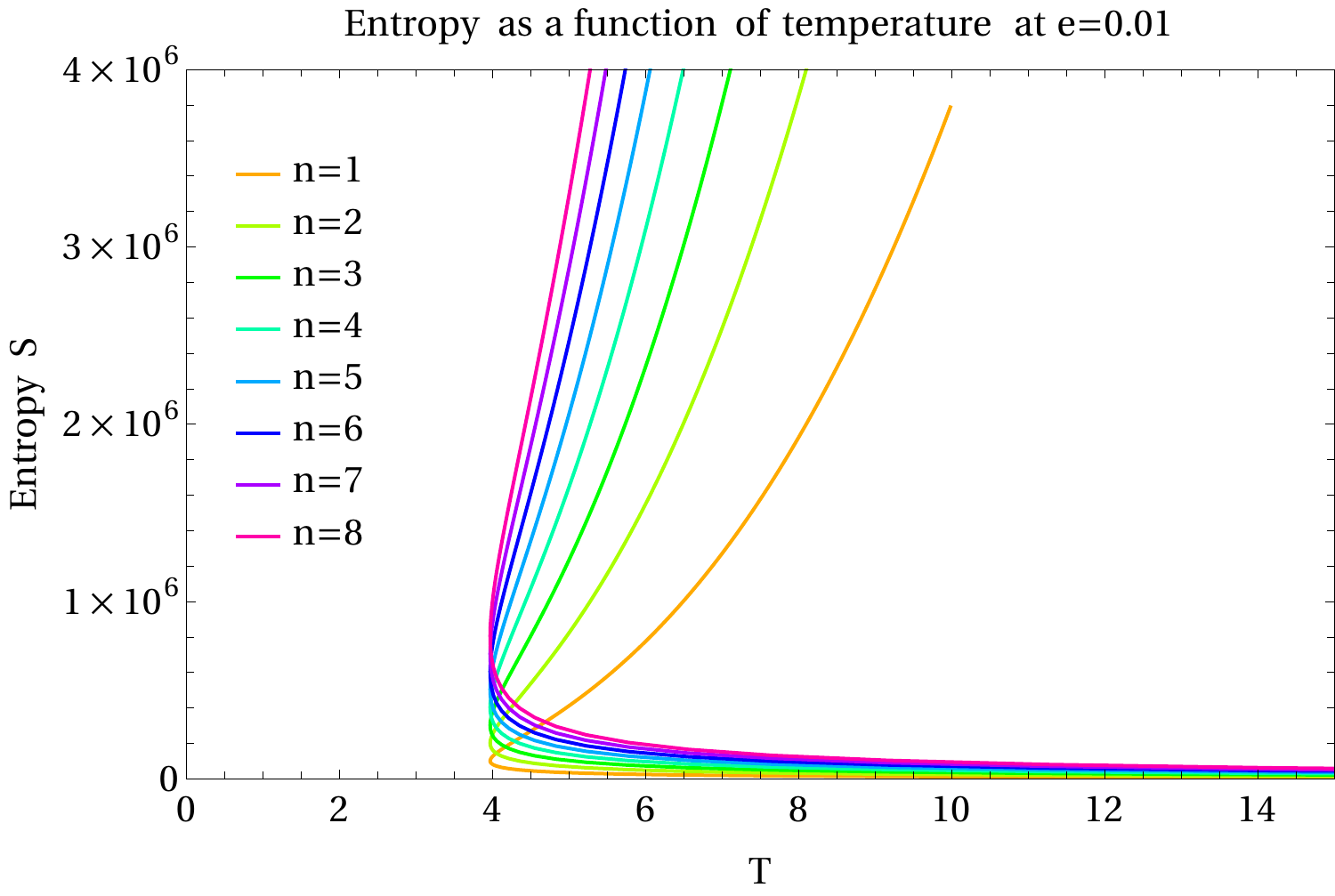} \hfill 
    \includegraphics[width=0.49\textwidth]{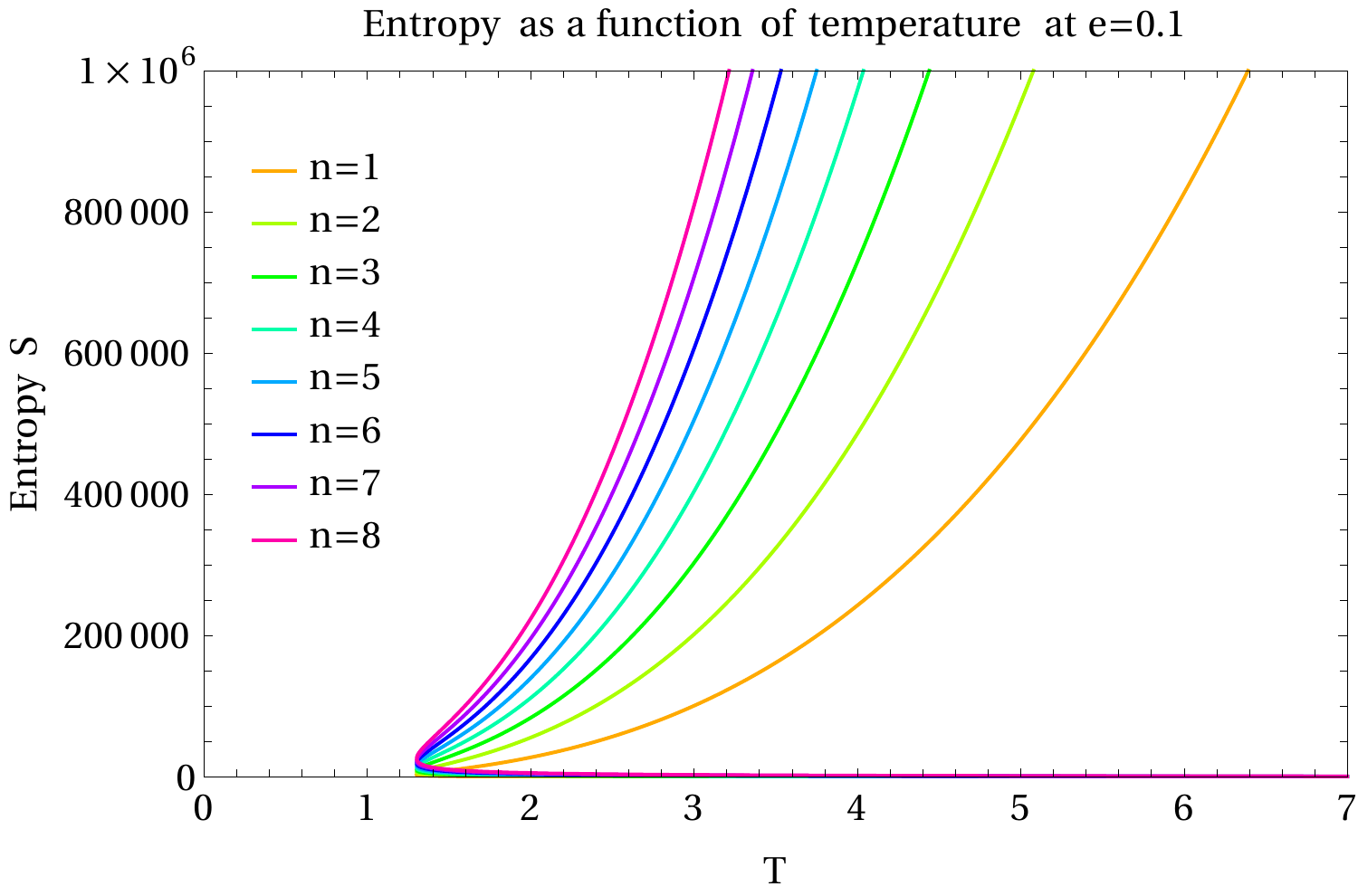} \hfill
    \includegraphics[width=0.49\textwidth]{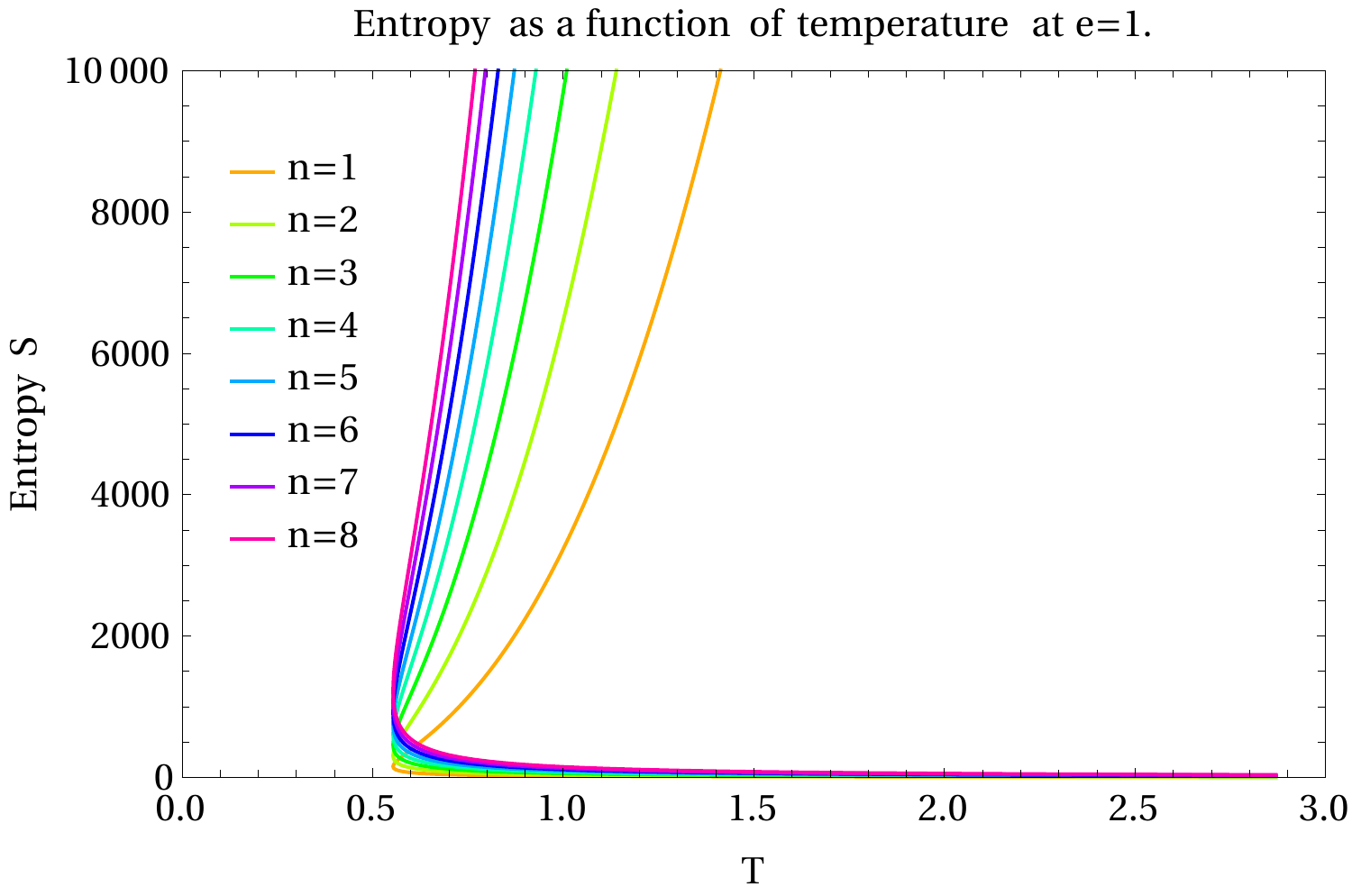}\hfill
    \includegraphics[width=0.49\textwidth]{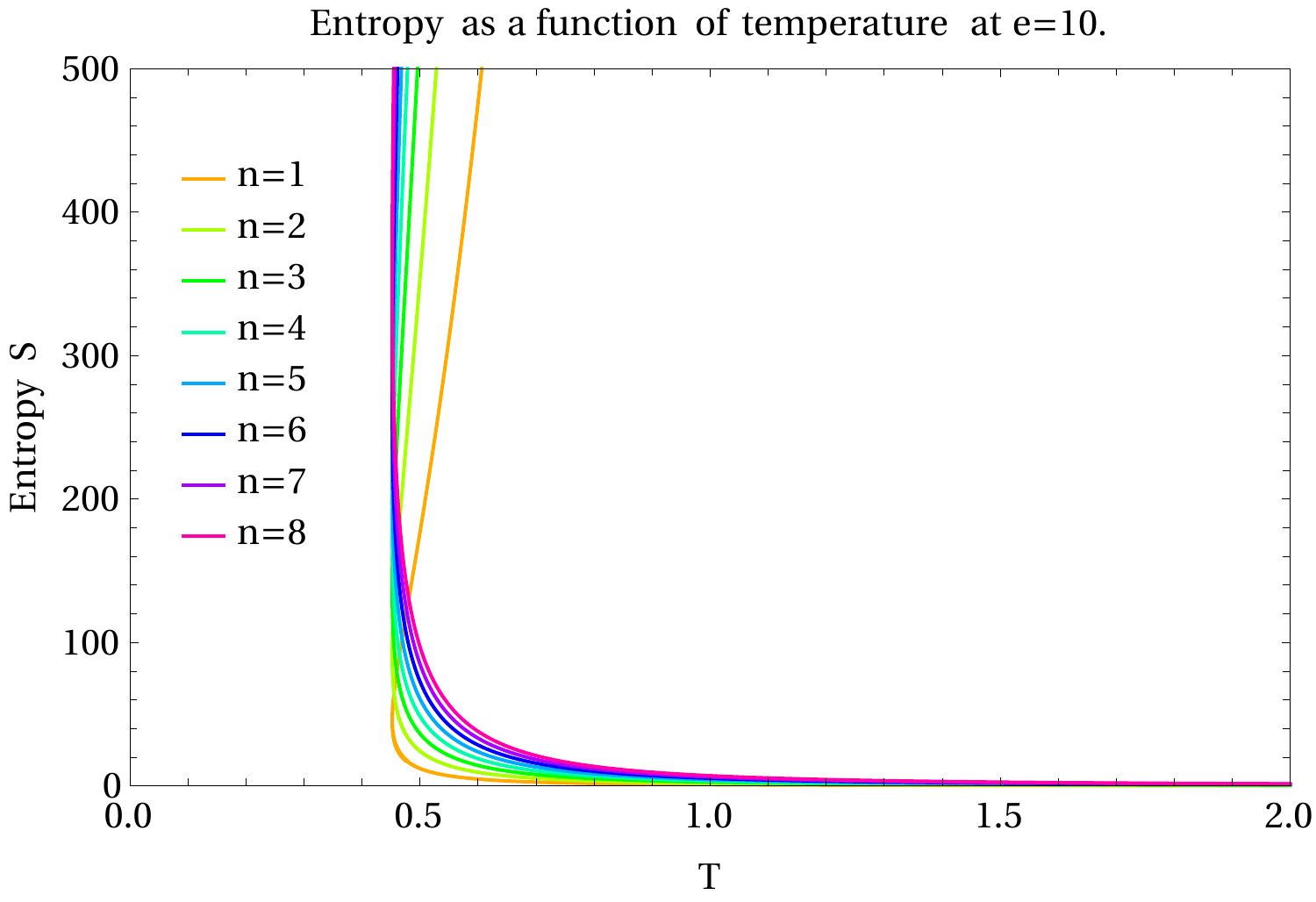}
    \end{center}
    \caption{
    \textit{Entropy density versus Temperature.} The entropy density displayed for various winding numbers as a function of the temperature of the solution.\label{fig:Top_S_v_T}}
    \end{figure}

\noindent\textbf{Entropy:} The entropy of the black hole is given by the area of the horizon,  which is found to be quantized 
\begin{equation}\label{eq:entropy}
    S =\frac{1}{4G_5}\int \exd^3 x \sqrt{\det(g_{ij})} =
    \frac{2\pi^2 n}{4\,G_5} r_h^3\, , \quad n\in \mathbb{N}  \, .
\end{equation}
This is in line with the fact that the energy of this solution gains a quantum of energy each time the winding number $n$ of the topological Skyrmion solution is increased. Figure~\ref{fig:Top_S_v_T} displays the dependence of the entropy on the temperature for various Skryme couplings. The solutions can be seen to be multi-valued. 

\noindent \textbf{Free energy: }According to the AdS/CFT dictionary the free energy of the dual $(3+1)$-dimensional field theory may be computed by means of the on-shell $(4+1)$-dimensional action by $\beta F = S_{ren}$. One may note that the action in eq. (\ref{eq:skyrmeAction}) is infinite and so must be renormalized by subtracting a counterterm.  The renormalized action is given by
\begin{equation}
 S_{ren}=  \lim_{\epsilon\rightarrow 0}\left(\int \exd^4 x\left(\int_{r_h}^{1/\epsilon}\exd r\sqrt{-g}\left(\frac{(R-2\Lambda)}{16\pi G_5}+\frac{\tr(F^2)}{16\pi \gamma^2}\right)   \right)+S_{ct}\right),
 \label{renaction}
\end{equation}
where $S_{ct}$ is the counter-term action. This action is given as the usual Gibbons-Hawking-York boundary term needed to make the variational problem well defined plus additional contributions needed to cancel both $1/\epsilon^n$ and $\log(\epsilon)$ divergences~\cite{Taylor:2000xw},
\begin{equation}
    S_{ct}=\frac{1}{8\pi G_5}\int\exd^4 x\sqrt{\gamma}\left( K -\frac{1}{2L}\left(2(1-d)-\frac{L^2}{d-2}R(\gamma)  \right)\right)+ \frac{L}{16\pi\gamma^2}\log(\epsilon)\int\exd^4 x\sqrt{\gamma_0}\tr(F_0^2)\, ,
\end{equation}
where $K$ is the trace of the extrinsic curvature, 
$\gamma$ is the induced metric on a constant $r=1/\epsilon$ hypersurface, $\gamma_0$ is the metric of the dual field theory and $F_0$ is the external field strength of the gauge field $A$ in the dual theory.

\subsection{Stability of our topological black hole}
Because gravitational interactions are attractive, any ensemble of matter is unstable against gravitational collapse. Even an ensemble of matter in thermal 
equilibrium has gravitational instabilities due to the fact that there is no way to shield against gravity. A typical gas compressed below a critical volume will collapse to undergo a phase transition, for example water vapor condensing to liquid water. 
Classical thermal gravitational systems also develop instabilities due to quantum effects. For example, quantum black holes can nucleate in flat spaces above a critical temperature, making such spaces unstable~\cite{Gross:1982cv}. However, black hole systems can be made thermodynamically stable with the addition of special boundary conditions or when  placed in spacetimes with a negative cosmological constant~\cite{Hawking:1975vcx,Gibbons:1976ue,York:1986yv}.  \\

\noindent\textbf{Stability conditions:} A system is stable against thermal fluctuations if  the  second  variation  of  the  entropy  with  respect  to  the  temperature  is negative, or  equivalently  if  the  heat  capacity  is  positive~\cite{Chamblin:1999hg}
\begin{equation}
   c_V=\left( \frac{\partial E}{\partial T}\right)_V\geq 0\, .
\end{equation} 
Likewise a system is stable against charge density fluctuations if  the  second  variation  of  the  entropy  with  respect  to  the  charge density  is negative, or  equivalently  if  the charge susceptibility, which is defined to be the variation of the chemical potential with respect to the charge density,  is  positive 
\begin{equation}
    \chi_{\rho} =\left( \frac{\partial \mu}{\partial \rho}\right)_V\geq 0\, .
\end{equation}
For the topological black hole to be discussed in section 3.2 with isospin ``charge'' the charge susceptibility $\chi_\rho$ is a tensor in isospin space, $ {(\chi_{\rho})}^a_b$, defined as
\begin{equation}
    {(\chi_{\rho})}^a_b=\frac{\partial Q^a}{\partial \mu^b}\, ,
\end{equation}
where $\rho$ here is a label and $a,b=1,2,3$ are the directions in isospin space. This then defines a relation between the charge density and chemical potential  $Q^a=\vev{J^a}= {(\chi_{\rho})}^a_b\mu^b$. Using this definition we can extract the charge susceptibility $(\chi_\rho)^a_b$ for which we require all elements to be positive definite for the solution to be stable against isospin charge fluctuations.  \\

\noindent\textbf{Conserved Charges:} The Komar integrals define conserved charges of the solutions~\cite{Balasubramanian:1999re}. Given a Killing vector $\xi$ and a time-like unit vector $u^{\mu}$ normal to a space-like hypersurface $\Sigma$ there is an associated conserved charge
\begin{equation}
  Q =-\int_{\Sigma}\sqrt{\text{det}(g_{(0)})}\xi_{i}u_{j}\vev{T^{ij}}.\label{eq:Conserved_Quant}
\end{equation}
For the thermal stability we consider a time-like Killing vector associated with time translation $\xi^t$, whose resulting conserved charge is the mass of the spacetime
\begin{equation}
    M =-\int_{\Sigma}\sqrt{\text{det}(g_{(0)})}\xi^t_{i}u_{j}\vev{T^{ij}}.\label{eq:Conserved_Charge_Mass}
\end{equation}

To study the thermal stability of the solution given in eq. (\ref{eq:Topo_non_trivial_Metric}) we first compute the Komar integral. Inserting the energy momentum tensor as found in~\cite{Cartwright:2020yoc} into this expression we find
\begin{equation}
    M=\frac{3 \pi  L^2 n}{32 G_5}+\frac{3 \pi\,  m_t\, n}{8 G_5}+\frac{3 \pi  n }{8 e^2 G_5}\log \left(\Lambda L\right) \, . \label{eq:Mass_equation}
\end{equation}
To compute the heat capacity we consider the following chain rule
\begin{align}
    c_V&=\left(\frac{\partial M}{\partial m_t}\right)\left(\frac{\partial m_t}{ \partial r_h}\right)\left(\frac{ \partial r_h}{ \partial T}\right) \\
    &=\frac{3 \pi^2  n r_h^3 \left(L^2 \left(2 e^2 r_h^2+1\right)+4 e^2 r_h^4\right)}{2 G_5 \left(L^2 \left(-2 e^2 r_h^2+3\right)+4 e^2 r_h^4\right)} \, . \label{eq:heat_cap}
\end{align}
It is clear from this expression that the heat capacity would only be negative if 
\begin{equation}
   L^2 \left(2 e^2 r_h^2+3\right)-4 e^2 r_h^4 <0 \, ,
\end{equation}
which places a constraint on the horizon radius for a particular value of the Skyrme coupling. The bound is saturated when
\begin{equation}
    r_h=r_h^*=\frac{1}{2}\left[L^2\pm \frac{1}{e}\sqrt{L^2\left(e^2L^2+12\right)}\right]^{1/2} \, ,\label{eq:max_rh}
\end{equation}
at this point the heat capacity diverges (as it switches its sign), i.e. $
\lim_{r\rightarrow r_h^{*-}} c_V \rightarrow \infty
$. Otherwise we find that our solutions have a positive heat capacity ($c_V>0$) provided that, $ r_h< r_h^* $. Furthermore there is a particular value implied by eq. (\ref{eq:max_rh}); for the mass of the black hole it is given in terms of the Skyrme coupling and the $AdS$ radius as
\begin{equation}
m_t= \frac{3 e^2 L^2+3 \sqrt{e^2 L^2 \left(e^2 L^2+12\right)}+4 \log \left(\frac{\sqrt{e^2 L^2 \left(e^2 L^2+12\right)}}{e^2}+L^2\right)+14-8 \log (2)}{8 e^2}  \, . 
\end{equation}
\begin{figure}[htbp]
    \centering
    \includegraphics[width=12cm]{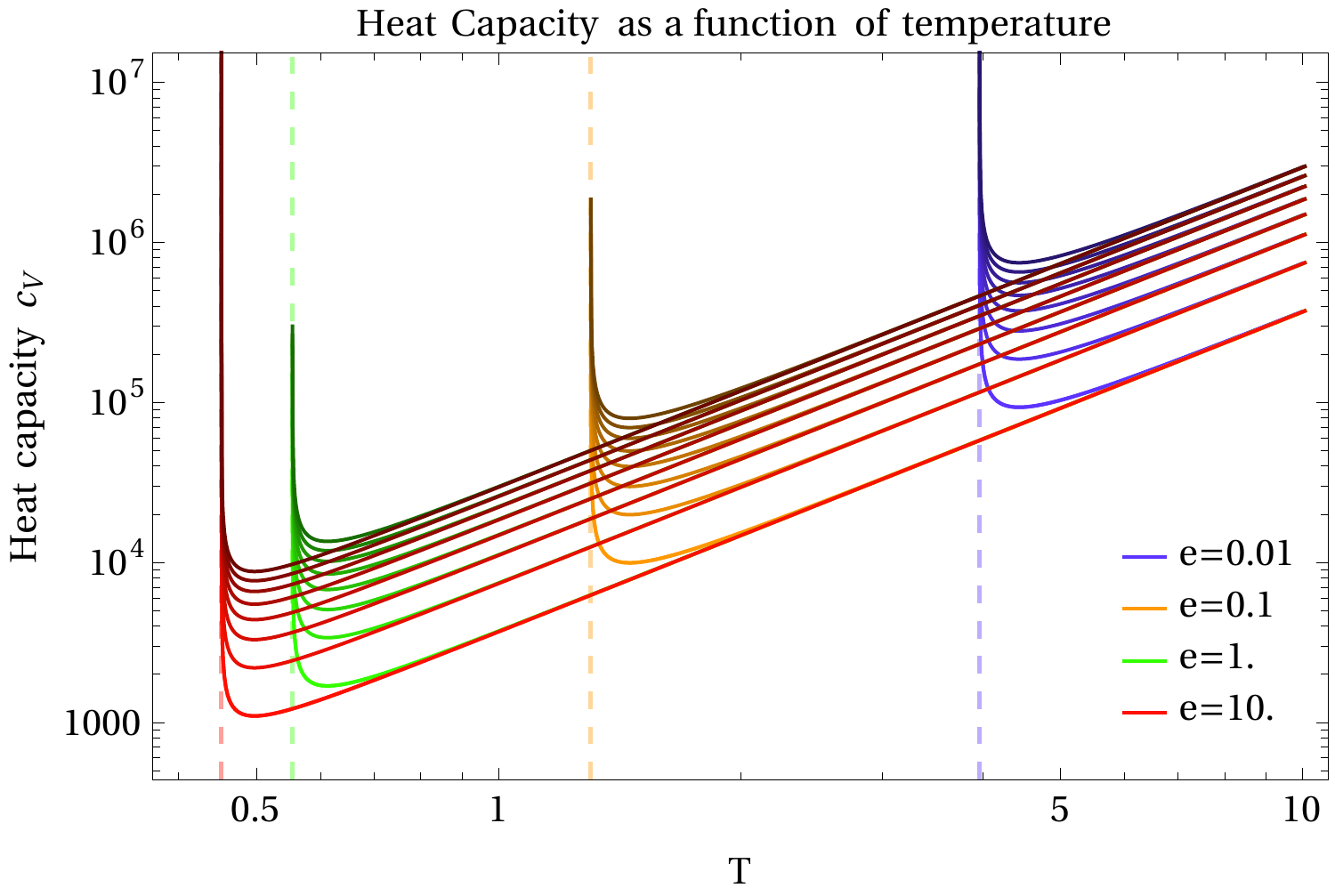}
    \caption{\textit{Heat capacity.} Here we display the heat capacity as calculated via eq. (\ref{eq:heat_cap}) as a function of temperature. The different colors represent different values of the Skyrme coupling $e$. We display this result for various winding numbers with the brightest color representing $n=1$ while the darkest color represents $n=8$. 
    \label{fig:heat_cap}}
\end{figure}

The charge density stability of the solution in eq. (\ref{eq:Topo_non_trivial_Metric}) is trivial for this solution to the Einstein equations. The values of all isospin charge densities is zero as can be seen in~\cite{Cartwright:2020yoc} hence $\chi_\rho=0$.

\subsection{Phase transition in our model}
\begin{figure}[htb]
\begin{center}
    \includegraphics[width=14cm]{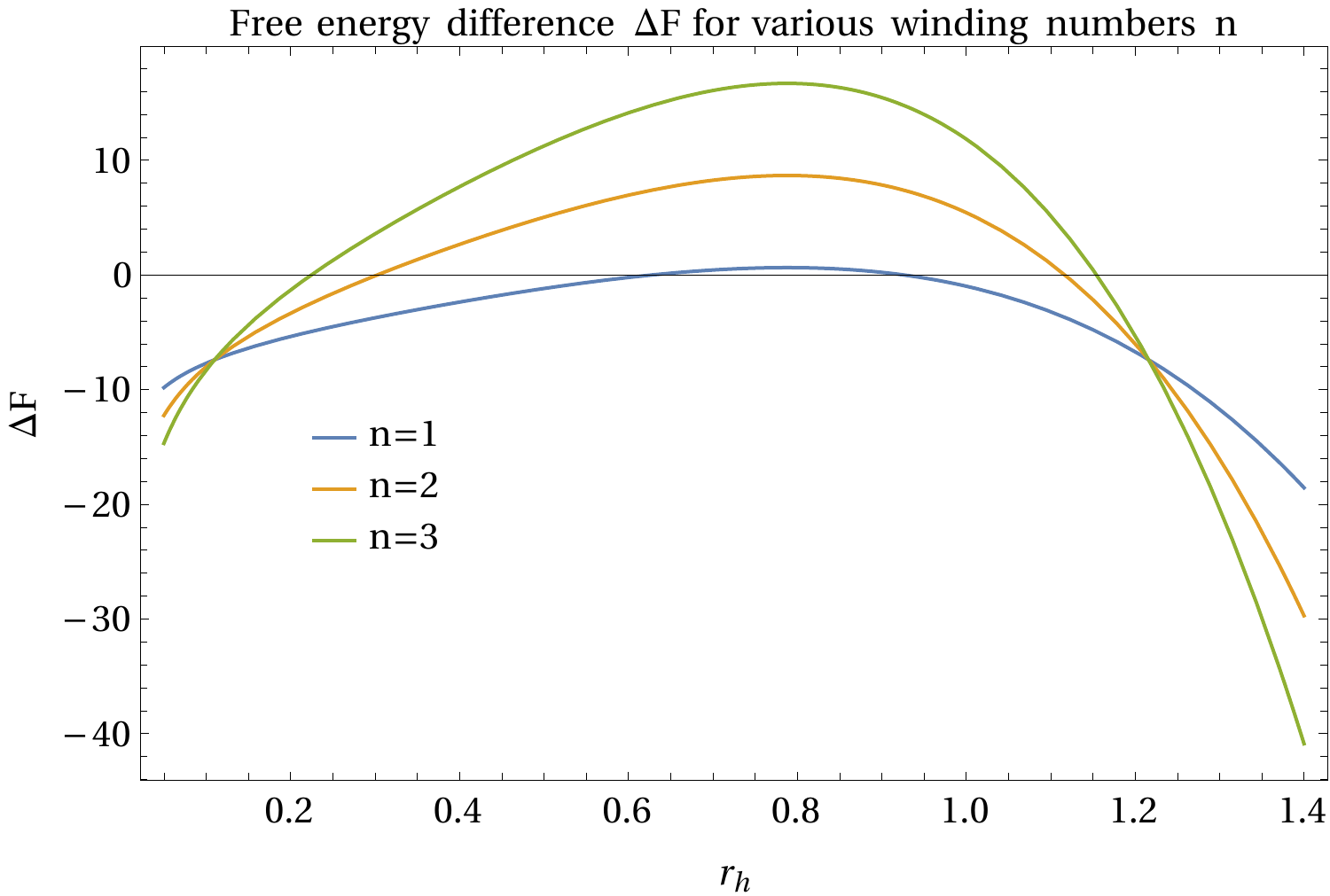}
    \end{center}
    \caption{
    \textit{The free energy difference plotted for winding numbers n = {\textcolor{blue}{1}}, {\textcolor{orange}{2}},     {\textcolor{green}{3}}.} For each value of the Skyrmion coupling parameter $e$ the free energy difference vanishes for two values of the horizon radius $r_h$.  Thus for each value of $e$ two phase transitions are possible, a transition to a large black hole or one to a smaller black hole. The origin of the ``focal points'', where the curves for all values of $n$ cross is described in the text.  Here, we chose $ e=\sqrt{10}$.
    \label{fig:Topologically_Non_Trivial_Free_Energy}}
\end{figure}

The well-known Hawking-Page transition~\cite{hawking1982} from thermal $AdS$ to an $AdS$ black hole spacetime occurs in our solution in the limit $e\rightarrow \infty$. In this limit the critical horizon value is $r_h=L$, which determines the critical temperature $T_c=3/(2\pi L)$. By computing the free energy associated with our topological Skyrme black hole solution and comparing it with that of thermal $AdS$ we discern the effect the Skyrmion has on the transition. 
The free energy difference between our topological $AdS$ black hole and thermal $AdS$ can be calculated from the renormalized action in eq.~\eqref{renaction}.  This difference is~\cite{Cartwright:2020yoc}\footnote{Note that eq.~(3.69) of the published version of~\cite{Cartwright:2020yoc} contains a typo. The factor $\beta$ on the right hand side should be removed. In addition the factors of $\tilde{e}$ should be $e$. However we note that in the choice of units used in that work $\kappa^2=8\pi G=1$, hence in figure 4. $\tilde{e}=e$. Also, the factor $1/\beta$ in figure 4 of~\cite{Cartwright:2020yoc} should be removed from the vertical axis label. }  
\begin{equation}\label{eq:DeltaFHawkingPage}
 \Delta F = F_{\text{BH}}-F_{\text{thermal}}= \frac{\pi ^2  n r_h^2 }{\kappa^2}\left(-\frac{e^2 L^4 L^2 \left(\frac{3}{n}-3\right)+4 e^2 r_h^4+4 L^2}{4 e^2 L^2 r_h^2}+\frac{3 \log (r_h)}{e^2 r_h^2}+1\right)\, ,
\end{equation}
where $F_{\text{thermal}}=3\pi^2 L^2 / (4\kappa^2 )$ and $\kappa^2=8\pi G_5$. 
If we take the limit $e\rightarrow\infty$, the horizon radius asymptotes to the horizon radius of the standard global $AdS_5$ black hole, $r_h \to r_h(e\to \infty)$, and leads to the free energy difference (for $n=1$),   
\begin{equation}
    \lim_{e\rightarrow\infty}\frac{\kappa^2 }{\pi ^2  r_h^2}\Delta F=\left(1-\frac{r_h^2}{L^2}\right) \, .
\end{equation}
This restores the standard Hawking-Page transition temperature defined by $r_h=L$. However, if $\infty>e>0$ and we consider e.g.~$n=1$, then the $\text{log}(r_h)$-term and the $1/(e^2 r_h^2)$-terms compete with $r_h^2/L^2$. 
The free energy difference~\eqref{eq:DeltaFHawkingPage} is displayed in figure~\ref{fig:Topologically_Non_Trivial_Free_Energy} as a function of the horizon radius.

\begin{figure}[htb]
\begin{center}
    \includegraphics[width=0.49\textwidth]{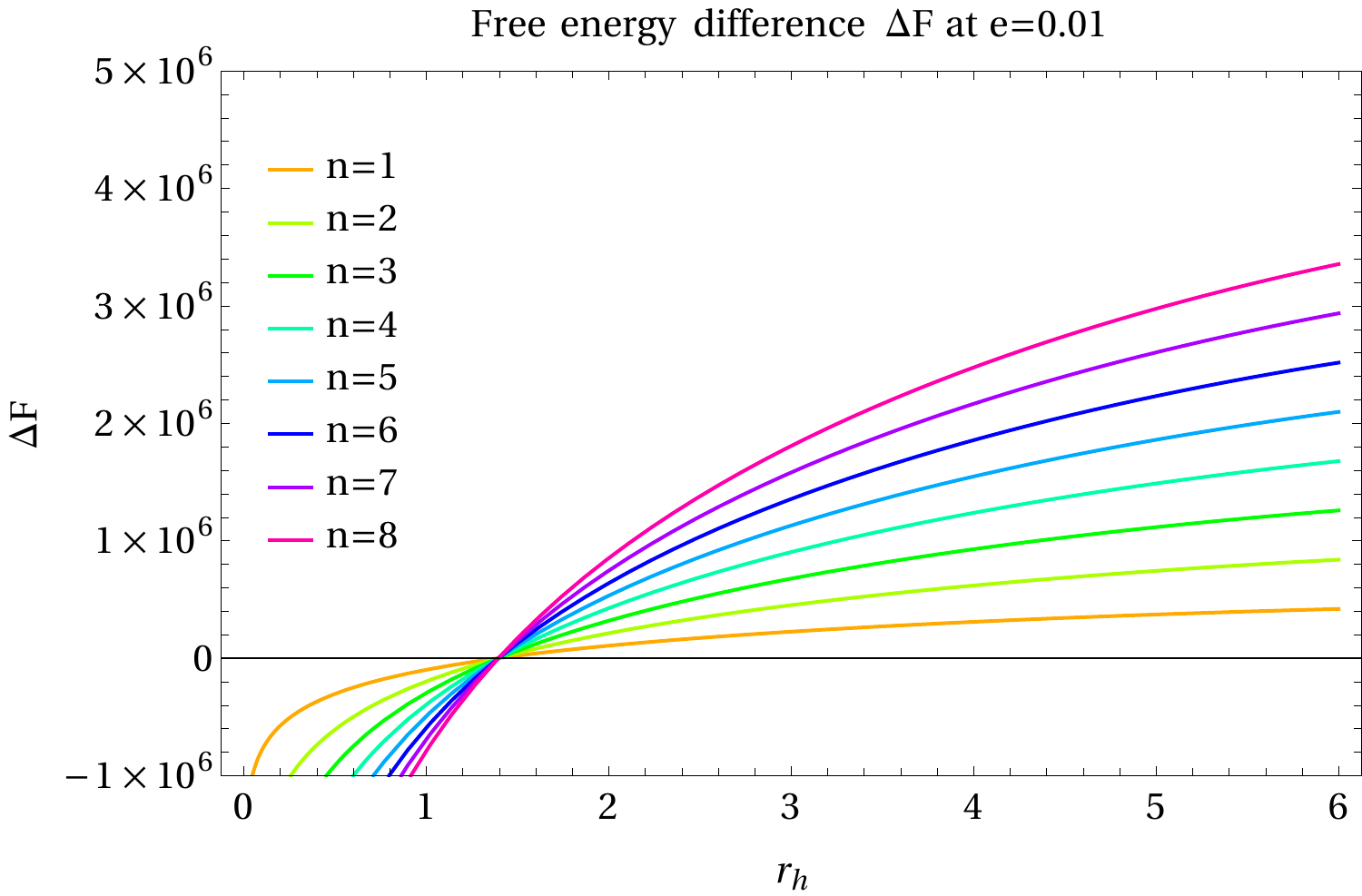} \hfill 
    \includegraphics[width=0.49\textwidth]{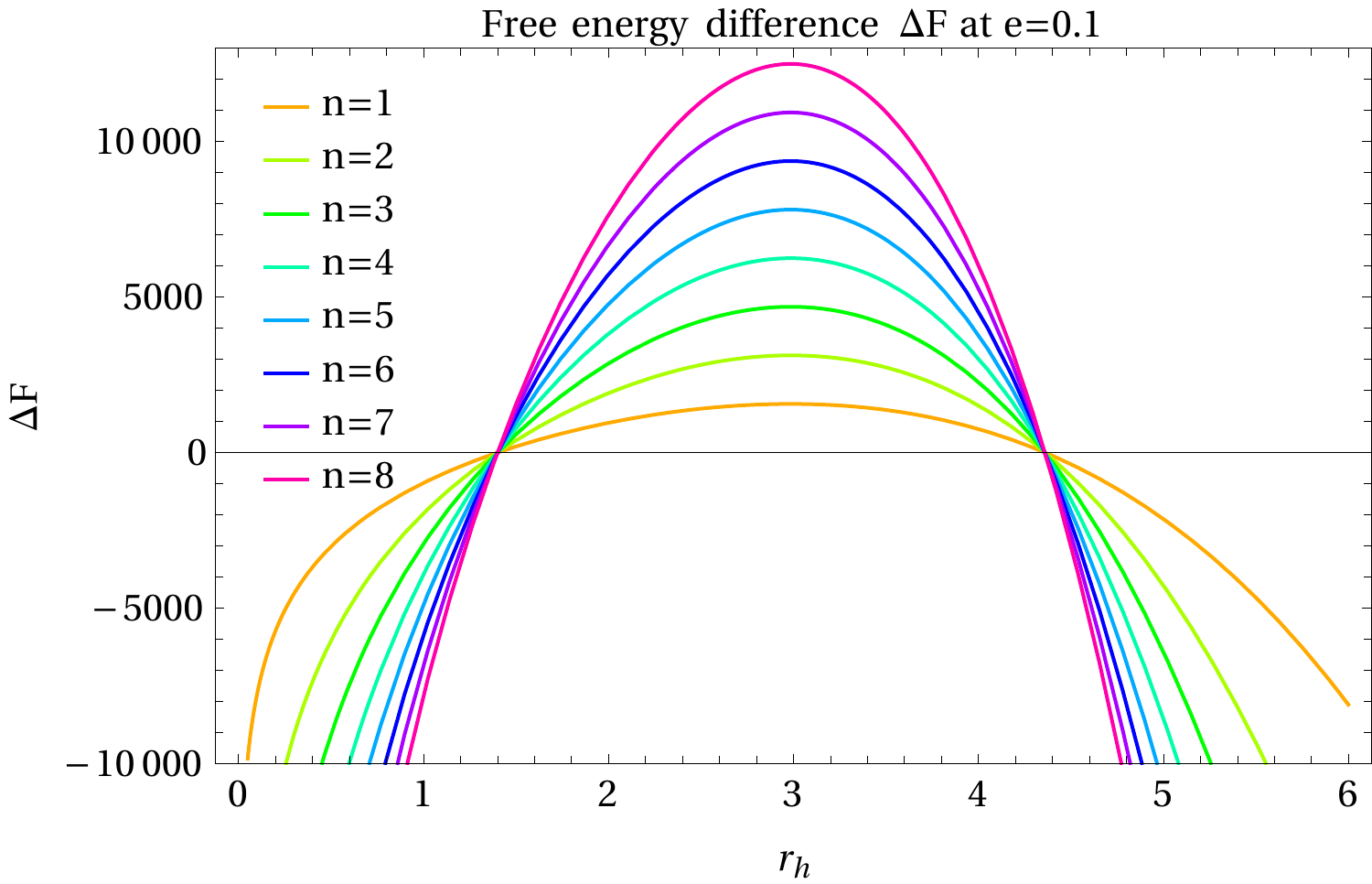} \hfill
    \includegraphics[width=0.49\textwidth]{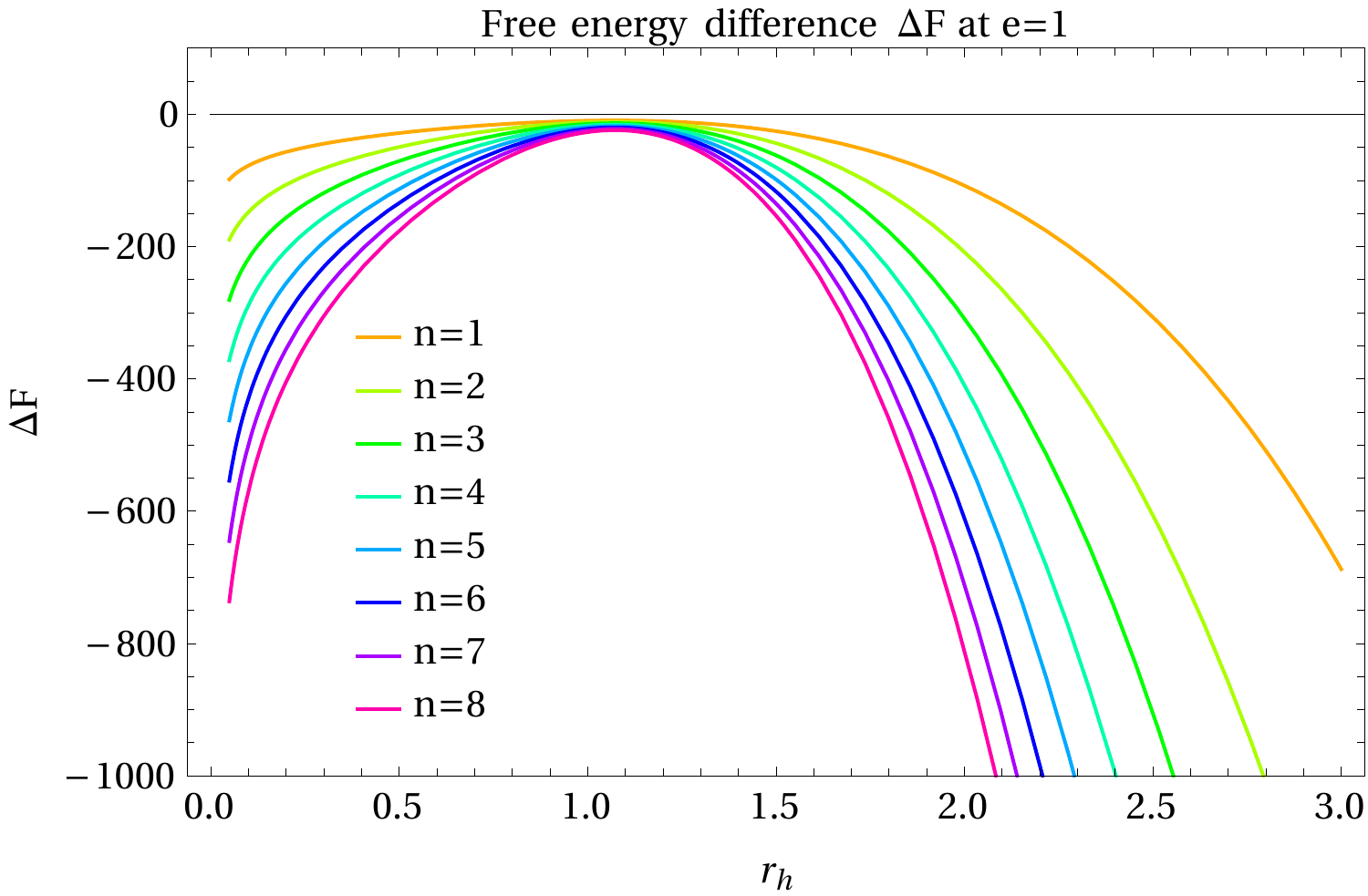}\hfill
    \includegraphics[width=0.49\textwidth]{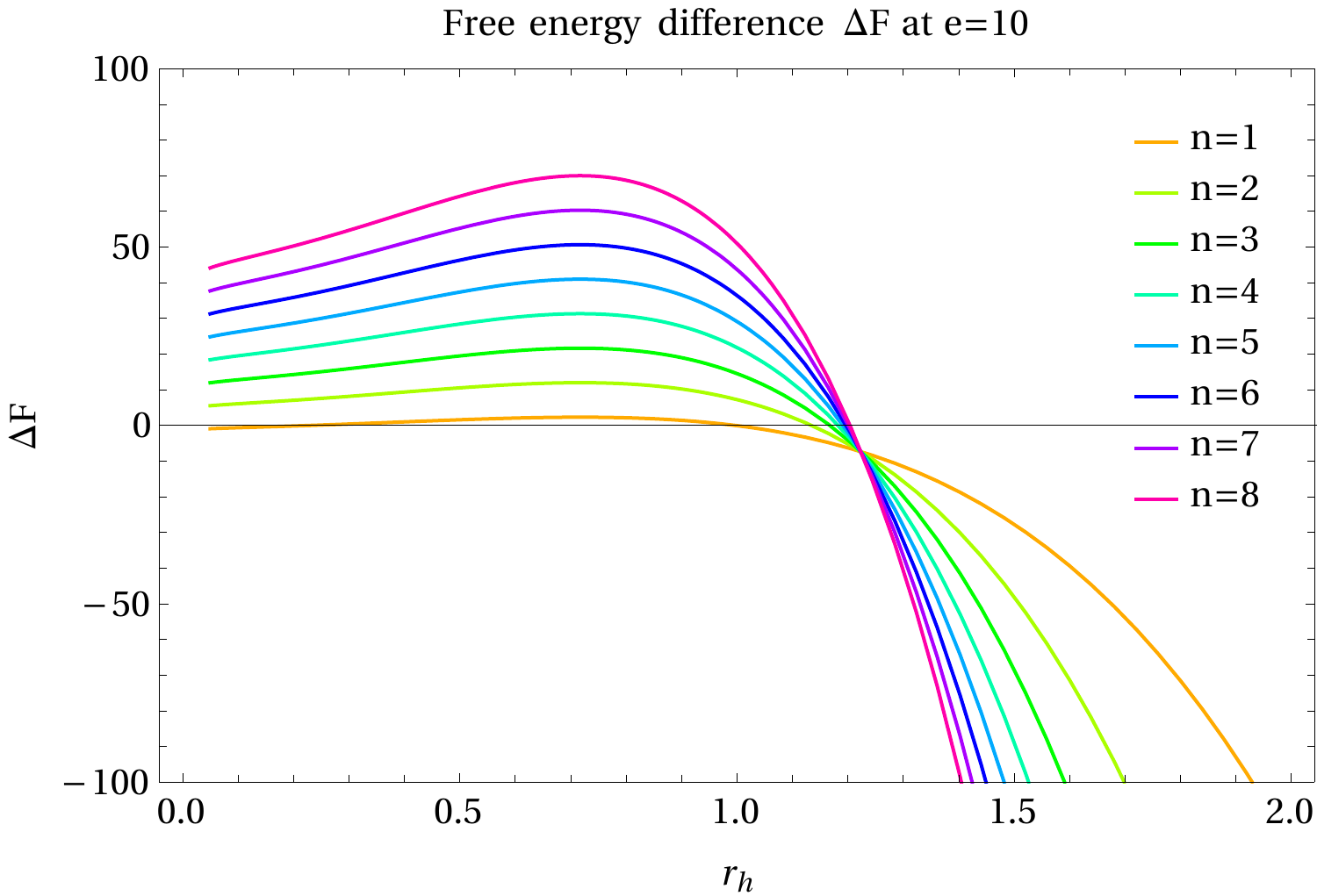}
    \end{center}
    \caption{
    \textit{Free energy difference versus horizon radius.} The free energy plotted for various winding numbers as a function of the horizon radius. In the top left image the free energy difference intersects the line $\Delta F=0$ again at a larger radius, $r_h$, outside the displayed plot range.  \label{fig:Top_F_v_rh}}
    \end{figure}
These results are generated at a value of the Skyrme coupling $ e=\sqrt{10}$. The lowest (blue) curve is virtually identical to the standard free energy for a thermal $AdS_5$ to global $AdS_5$ black hole transition at a critical temperature fixed by $r_h=L$. Increasing the winding number of our solution, this critical horizon radius increases by a $\Delta r_h$ which is a discrete function of $n$, shifting the Hawking-Page phase transition. This trend continues for larger $n>2$ as well. Below the transition, the configurations with larger $n$ have a larger free energy. However, there exists a ``focal point'' at a certain $r_h$ where all the free energies intersect. That means there is no energetically favored topology at that horizon radius (or corresponding temperature). Above that horizon radius, larger winding numbers are energetically preferred. As can be seen from \eqref{eq:DeltaFHawkingPage}, the contribution from the thermal $AdS$ free energy, $F_{thermal}$, (the $3/n$-term), which is multiplied by a factor of $n$, is independent of the winding number $n$.  The remaining terms in the expression are proportional to $n$, which can be factored out, leaving an expression which is a polynomial in $r_h$.  The polynomial vanishes at $ r_h\,=\,0.113 $ and $r_h\,=\, 0.22$ for $e\,=\,\sqrt{10}$ and $\kappa\,=\,1$, leaving $F_{thermal}$ as the only contribution to $\Delta F$. Thus the parameters of the Skyrmion field can be chosen such that the free energy of the topological black hole is zero for all winding numbers.
In other words, in figure~\ref{fig:Topologically_Non_Trivial_Free_Energy} at $r_h\,=\,0.113$ and $r_h=1.22$ there is only one contribution to $\Delta F$, which is independent of $n$, hence the curves for different $n$ intersect at the focal points at those horizon values. 

\begin{figure}[htb]
\begin{center}
    \includegraphics[width=0.49\textwidth]{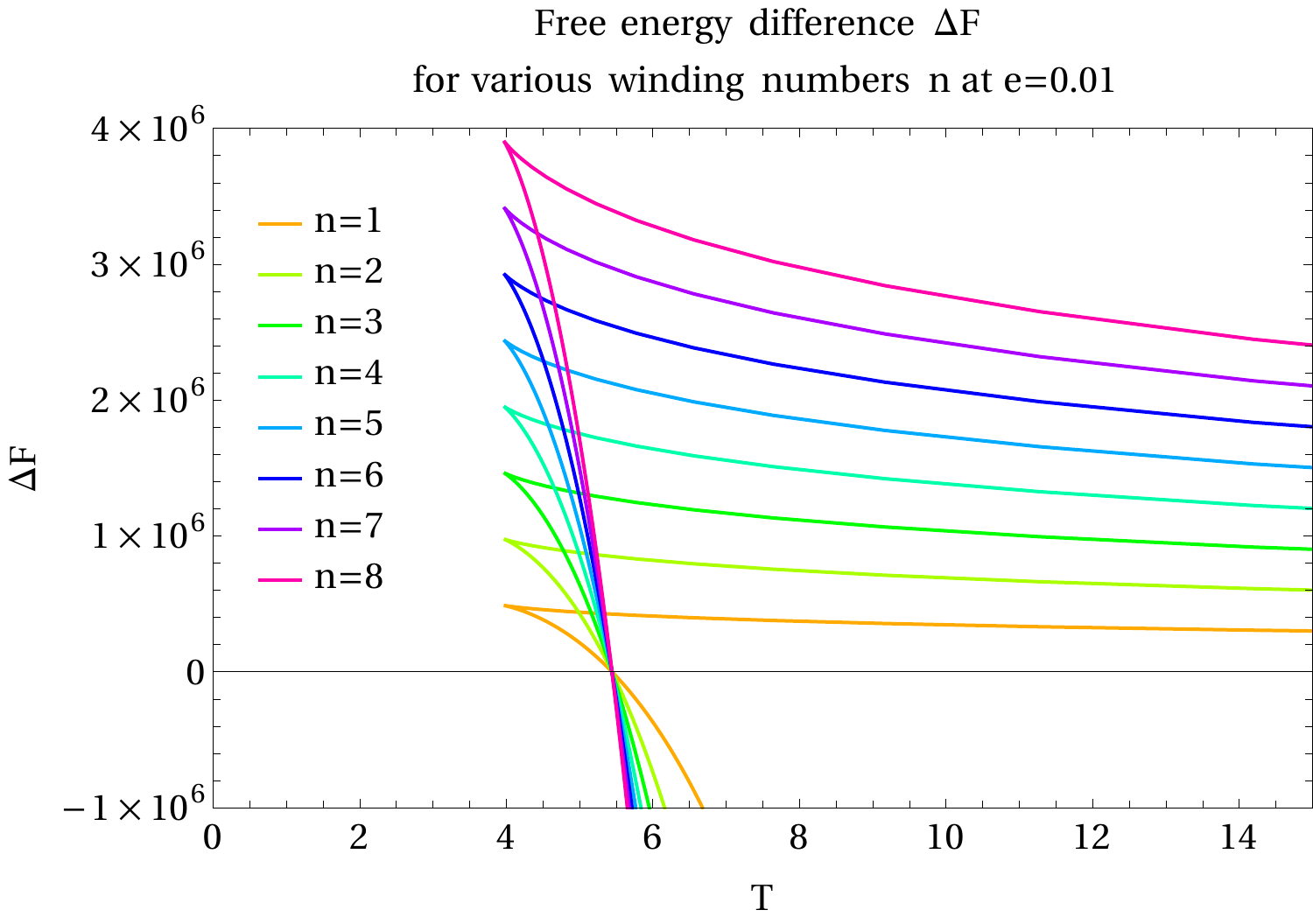} \hfill 
    \includegraphics[width=0.49\textwidth]{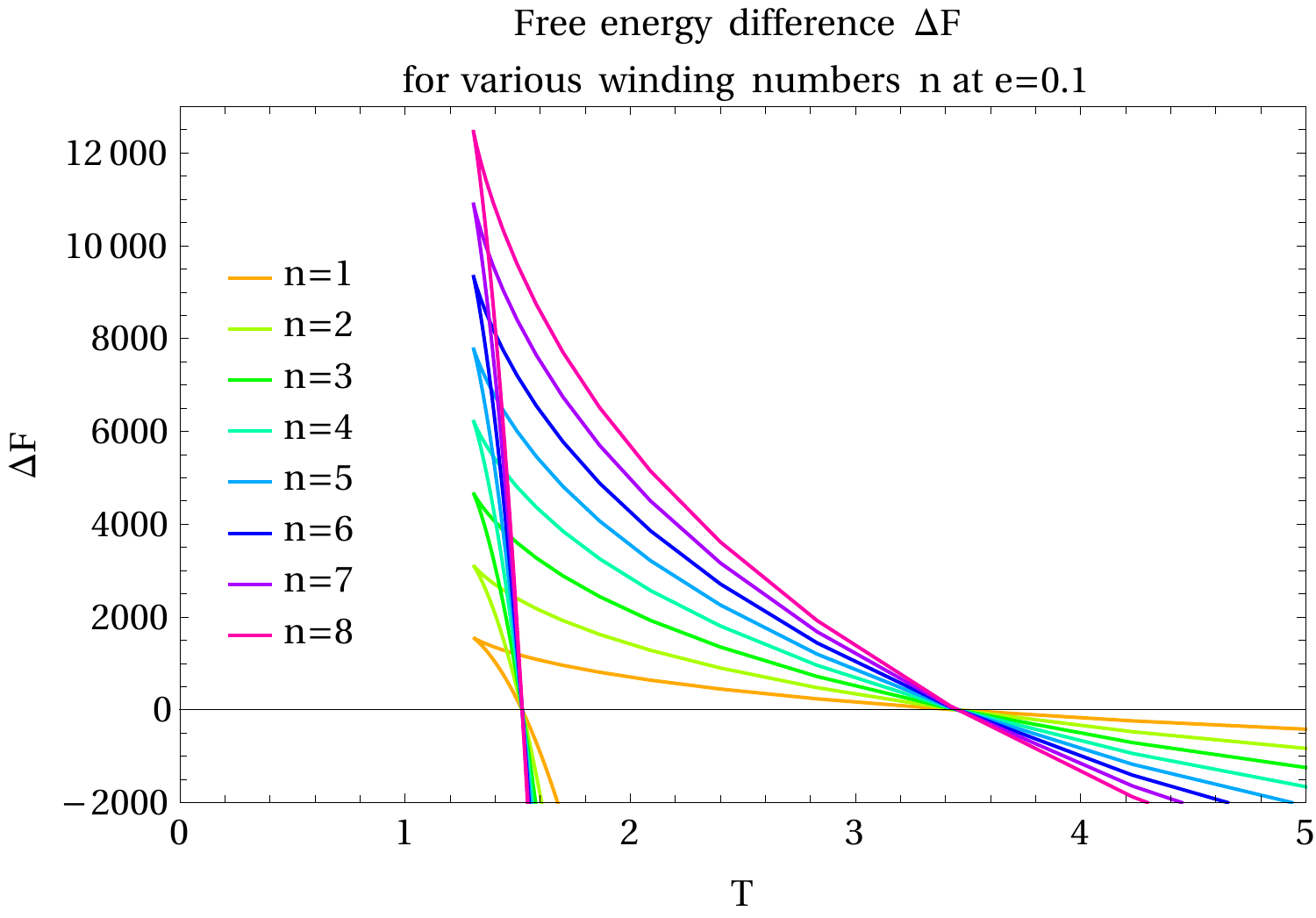} \hfill
    \includegraphics[width=0.49\textwidth]{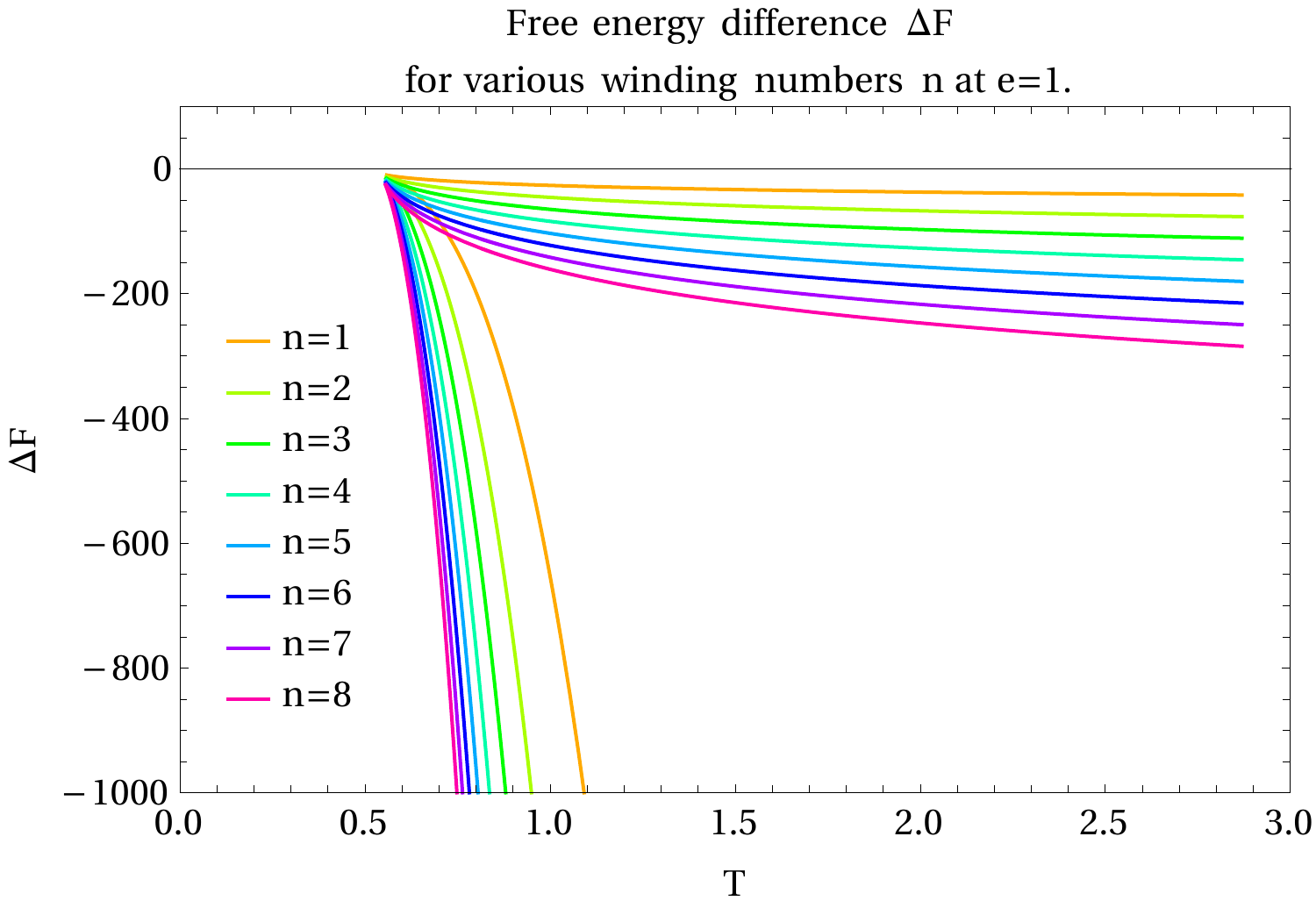}\hfill
    \includegraphics[width=0.49\textwidth]{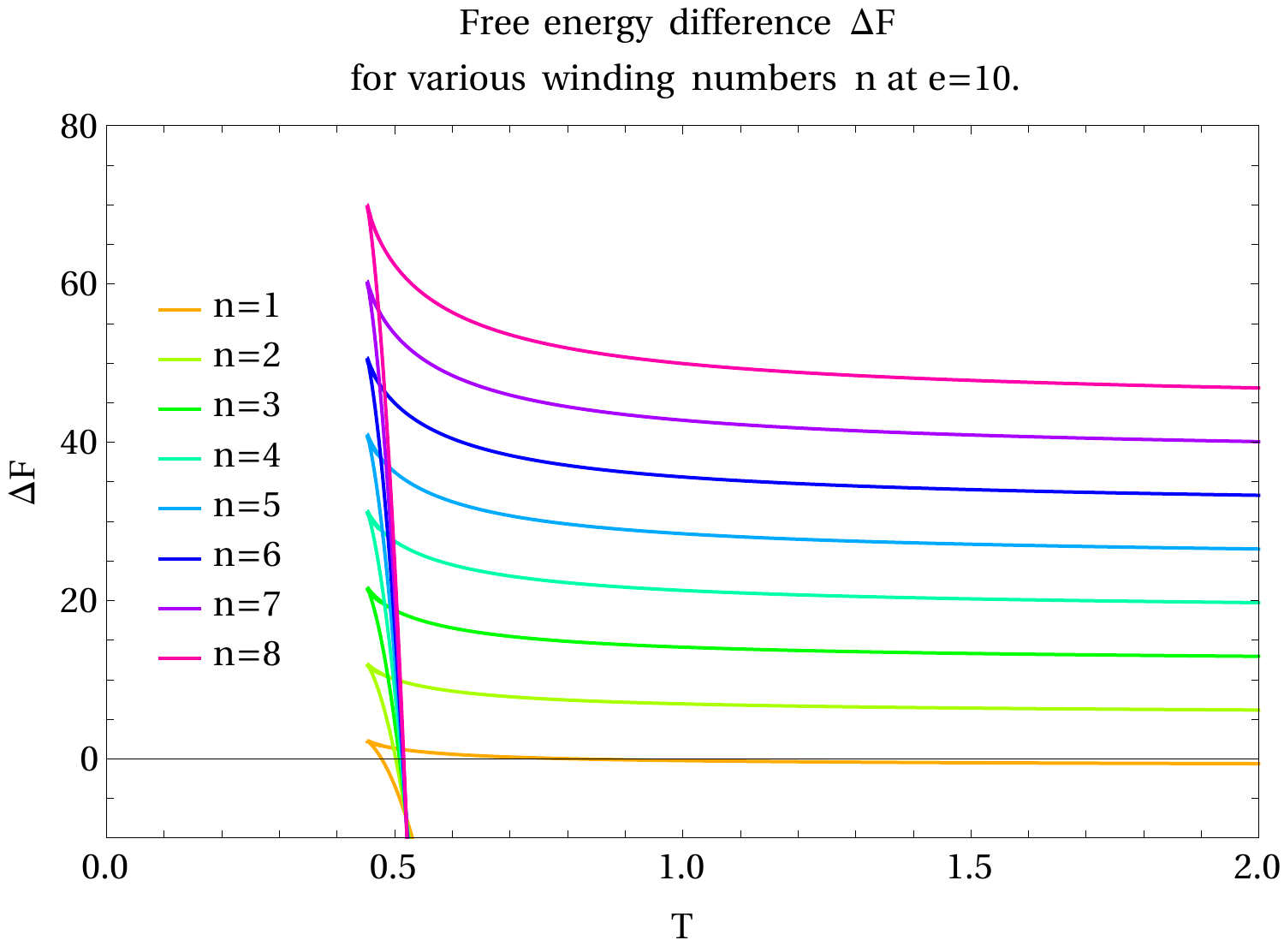}
    \end{center}
    \caption{
    \textit{Free energy versus temperature.} The free energy plotted for various winding numbers as a function of temperature. In the top panel all curves intersect the $\Delta F=0$ axis twice, once at a lower critical temperature $T_{c,L}$ and once at a higher one $T_{c,H}$. For the top left plot this intersection at $T_{c,H}$ is outside the displayed plot range. 
    \label{fig:Top_F_v_T}}
\end{figure}

In figure~\ref{fig:Top_F_v_rh}, the difference $\Delta F$  between the free energy corresponding to the topological black hole and that corresponding to thermal AdS, according to eq.~\eqref{eq:DeltaFHawkingPage}, is shown as a function of the horizon radius $r_h$. 
Similarly, in figure~\ref{fig:Top_F_v_T}, this quantity is displayed as a function of the black hole temperature. 
Just like in figure~\ref{fig:Topologically_Non_Trivial_Free_Energy}, increasing the temperature, for some combinations of $n$ and $e$ we observe two transitions: first a confinement and then a subsequent deconfinement transition.  
As the free energy of the dual field theory in figure~\ref{fig:Top_F_v_rh} suggests, at high temperatures corresponding to large values of $r_h$, the free energy difference is negative $\Delta F<0$, indicating that the deconfined phase (topological black hole) is preferred. 
Decreasing the temperature, for any given $n$ we find a confinement transition to  (H - P transition to thermal AdS) at a transition radius corresponding to the H - P transition temperature. For large values of the Skyrme coupling (lower right plot in figure~\ref{fig:Top_F_v_rh}), most values of $n$ produces only this single transition analogous to the original confinement transition dual to the H - P transition~\cite{Witten:1998zw}, with the latter case reproduced exactly at $e\to \infty$ for $n=1$. For the large but finite value, $e=10$, shown in the lower right plot of figure~\ref{fig:Top_F_v_rh}, the case $n=1$ undergoes a second transition at a smaller radius $r_h$. Since this corresponds to a transition from the confined to the deconfined phase, we refer to this second H - P transition as the {\it deconfinement transition}. For small values of $e\le 0.1$, the top right plot in figure~\ref{fig:Top_F_v_rh} shows that with stronger influence of the Skyrme sector (smaller $e$) all the windings $n\ge 1$ undergo two transitions, one forming a large black hole and one forming a small black hole.  This is still true for the top left plot in figure~\ref{fig:Top_F_v_rh}, the first (large black hole) transition is pushed to very large values of $r_h$ beyond the plot range. Furthermore, there exists an intermediate regime around $e\approx 1.$, where no H - P transition occurs, and the theory remains in its deconfined phase for all radii as $\Delta F <0 \forall T$ at all $n$. \\
In order to understand this phase transition in field theory terms, we consider the free energy as a function of temperature~\ref{fig:Top_F_v_T}. Increasing the temperature from $T=0$, we are first located in the confined phase (thermal AdS solution). No black hole solution exists there. For any given winding number, $n$, there is a minimal temperature $T_{\text{min}}$, at which the black hole solutions start to exist. In the top left panel of figure~\ref{fig:Top_F_v_T}, we find that for all values of $n\ge 1$ the black hole phase is not preferred at that $T_{\text{min}}$, because $\Delta F>0$. Increasing the temperature further, we can follow one of two branches and find it intersects $\Delta F=0$ at a low critical transition temperature, $T_{c,L}$. This is the new transition facilitated by the coupling to the Skyrmion (dual to the topological gauge field configuration in the field theory). Following the other branch, one arrives at a version of the original H - P transition at a high critical temperature $T_{c,H}$. Note that for the top panel cases $e=0.01$ and $0.1$, the critical temperatures $T_{c,L/H}$ are virtually independent from $n$. In contrast to that, the bottom right plot in figure~\ref{fig:Top_F_v_T} shows that the critical temperatures in the case $e=10$ increase with increasing $n$.

\section{Holographic interpretation}
\label{sec:holo}
In this section, we remind the reader of an interpretation of our solution from section~\ref{sec:gravity} in the holographically dual field theory as given in~\cite{Cartwright:2020yoc}, where this topic is expounded upon in much further detail. In summary, the Skyrme coupling and Skyrme field correspond to $SU(2)$-gauge fields and their coupling to $\mathcal{N}=4$ SYM theory, respectively.  

\noindent\textbf{Skyrmions as merons}: In~\cite{Cartwright:2020yoc} we discussed that one can consider Skyrmion solutions as being equivalent to specific $SU(2)$ gauge field configurations known as merons~\cite{Canfora:2013osa,Ayon-Beato:2015eca,Canfora:2017yio,Canfora:2018ppu,Ayon-Beato:2019tvu,Ipinza:2020xgc}, classical topological soliton solutions~\cite{RevModPhys.51.461}. Merons were originally considered in Euclidean space-time and interpreted as {\it half-instantons}. By definition $U^{-1}\partial U$ is a pure gauge solution, hence merons with $A=\frac{1}{2} K =\frac{1}{2} U^{-1}\partial U$ are {\it half of pure gauge solutions}. In particular the action of this meron theory is given by
\begin{equation}\label{eq:meronAction}
 S_{\text{meron}} = \int d^5x \sqrt{-g}
 \left (
 \frac{1}{16\pi G} (R-2\Lambda) 
 + \frac{1}{16\pi \gamma^2} \text{Tr}[
 F_{\mu\nu} F^{\mu\nu} - 2 m^2 A^\mu A_\nu
 ]
 \right) \, ,
\end{equation}
with the Yang-Mills coupling constant $\gamma$, the cosmological constant $\Lambda=-6/L^2$ with $L$ the $AdS$ radius and the Proca mass $m$~\cite{Kunimasa:1967GT,Shizuya:1975ek}. If we identify  particular values of the parameter $\lambda$, the meron and the Skyrmion theory have the same solutions. The exact mapping is given by,
\begin{equation}
    A_\mu=\lambda K_\mu \, , \quad 
    F_{\mu\nu} = \lambda (\lambda-1) [K_\mu,K_\nu] \, , \quad 
    \frac{m^2 \lambda^2}{\pi \gamma^2} = \frac{f_\pi^2}{2}\, , \quad 
    \frac{\lambda^2(\lambda-1)^2}{\pi \gamma^2} = 
    \frac{1}{2 \tilde{e}^2}\, ,\label{eq:identify}
\end{equation}
where it is crucially important that $|\lambda|=1/2$ and $f_\pi$ is the coupling constant multiplying the kinetic Skyrmion term in the action eq.~\eqref{eq:skyrmeAction}. When evaluated at vanishing Proca mass, $m=0$, the meron action~\eqref{eq:meronAction} coincides with the Einstein-Yang-Mills action with an $SU(2)$ gauge field, $A_\mu$. This represents a consistent truncation of the bosonic part of minimal gauged type IIB supergravity in five dimensions. While our analysis does not require the mass term, such a term can be generated within type IIB supergravity through spontaneous symmetry breaking in the bulk~\cite{Klebanov:2002gr}. In other words, the massless meron theory corresponds to $\mathcal{N}=4$ Super-Yang-Mills theory~(SYM) in flat Minkowski space coupled to an external $SU(2)$ gauge field $F$ associated with an $SU(2)$ subgroup of the $SU(4)$ R-symmetry of the $\mathcal{N}=4$ SYM theory, see e.g.~\cite{Son:2006em,Behrndt:1998jd,Cvetic:1999ne,Gubser:1998jb,Chamblin:1999tk}.\\

In the AdS/CFT correspondence, black branes/holes are generally regarded to be dual to a state of a strongly coupled quantum many-body system at non-zero temperature. One may imagine the quark-gluon-plasma generated in heavy ion collisions, or a strongly-correlated electron system. Meanwhile, the Skyrme hair of the branes is dual to the gauge field sourcing the $SU(2)$ current $J^\mu$. In the case at hand the dual current operator is restricted in that it is dual only to merons, pure gauge solutions. 
In~\cite{Cartwright:2020yoc} the topological configurations of the associated $SU(2)$ gauge field, $A_m$, on the field theory side were determined from the near-boundary exponents of the normalizable and non-normalizable modes of our meron gauge field $A_\mu$, in the gravity bulk given by the solution in eq.~(\ref{eq:top_Ansatz}) and eq.~(\ref{eq:Topo_non_trivial_Metric}). 
These topological configurations are characterized by the integer winding number  
\begin{equation} 
n
= - i \frac{e^3}{12\pi^2} \oint\limits_{S^3}\exd^3 x \, \hat n_i \epsilon^{ijkm}  \text{tr}(A_j A_k A_m) \, .
\end{equation}

In summary, the two key features of our holographic model are confinement and the topological winding number $n$. Various systems share these key features with our model, for example there is a notion of confinement on spin chains which are also topologically non-trivial~\cite{Bera:2017,Lake:2009,Vestergren:2005in}. As we alluded to in section~\ref{sec:intro}, application of our model to such condensed matter systems provides an exciting prospect for the future.

\section{Relation of phase transitions in our model to phase transitions in other physical systems}
\label{sec:transitions}
%
\subsection{Bekenstein's atomic black hole model}
In \cite{Bekenstein:1995ju,Bekenstein:1997bt} the spectroscopy of a four-dimensional, quantum black hole was obtained by treating the area of the horizon of a black hole as an adiabatic invariant. An analogy was made between non-extremal quantum black holes and the allowed energy levels in an atom by assuming that the area of a black hole is quantized. The area of a black hole increases or decreases only by absorbing or emitting area quanta.  In this approach the area of the horizon is found to be linearly proportional to an integer $N$, which implies that the mass of the black hole is proportional to the square root of $N$
\be M \,\propto\, \sqrt{N} \, .
\ee 

This proportionality leads to a partition function which is not well-defined if the integer $N$ has the range $[0, \infty]$.  The partition function for this system is 
\be 
Z[T ] = \sum_{N = 0}^{\infty}\,k^N\,\exp^{-\gamma\,\frac{\sqrt{N}}{T}}\, ,
\ee
where $\gamma\,=\,\sqrt{\alpha/(16\,\pi)}$, and $\alpha\,=\,4\,\ln(k)\, , \, k\,=\, 2,4,...$ 
A well-defined partition function can be obtained by putting the system in a box of radius $r$ \cite{Stephens:2001sd}, which restricts the range of $N$ to $[0, N_{max}]$ where 
\be 
N_{max}\,=\,\frac{r^2}{4 \gamma^2}\, .
\label{Nmax}
\ee
The partition function is now
\be 
Z(T)\,=\,\sum_{N = 0}^{N_{max}}k^N\,e^{-\gamma\sqrt{N}/T}\, ,
\ee
where $k$ is the degeneracy of each state.  The average energy and entropy calculated from this partition function are qualitatively similar to the corresponding quantities which characterize the nucleation of black holes.

\begin{figure}[htb]
\begin{subfigure}[b]{0.5\textwidth}
    \includegraphics[width=7cm]{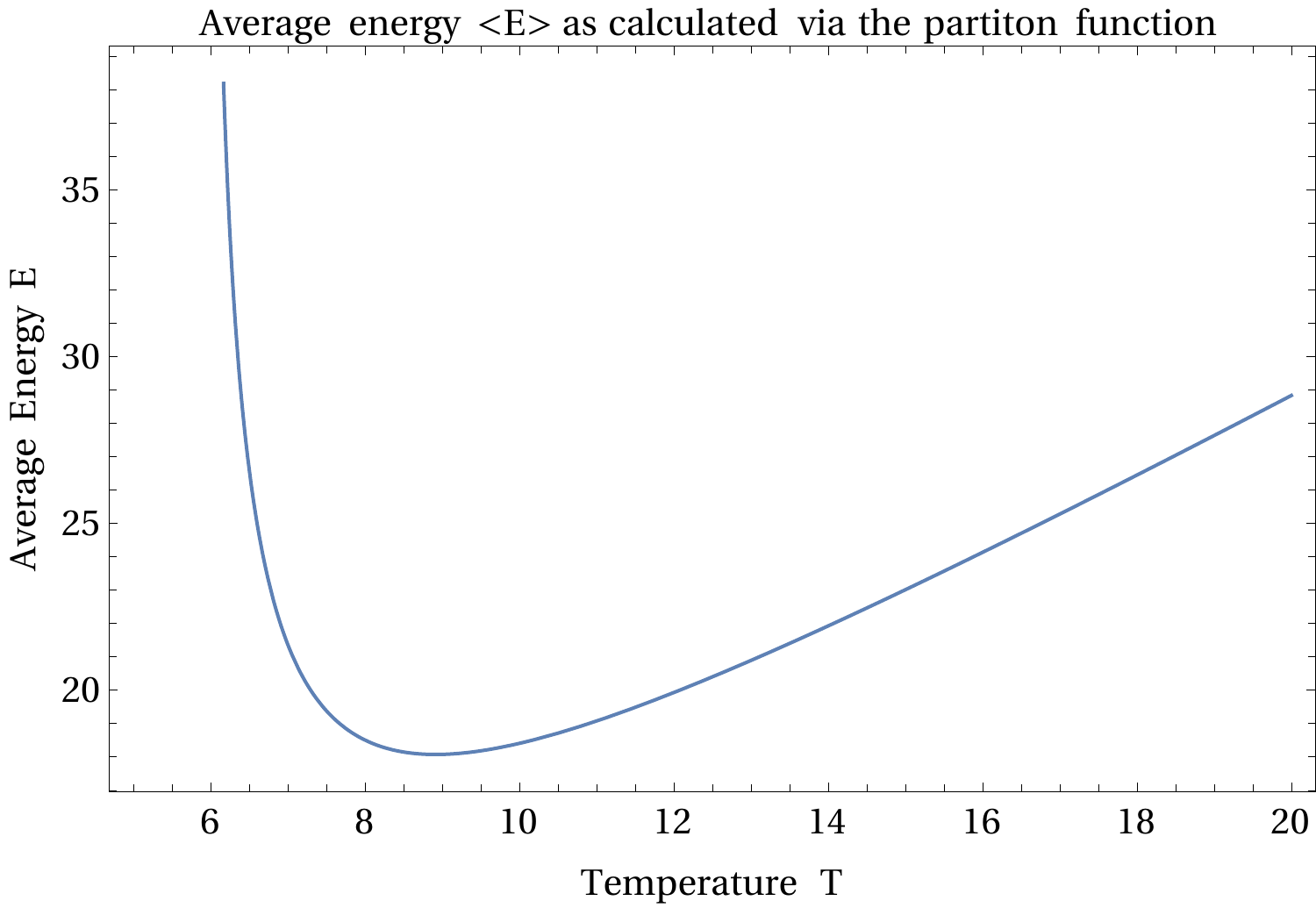}
    \end{subfigure}
    \begin{subfigure}[b]{0.5\textwidth}
    \includegraphics[width=7cm]{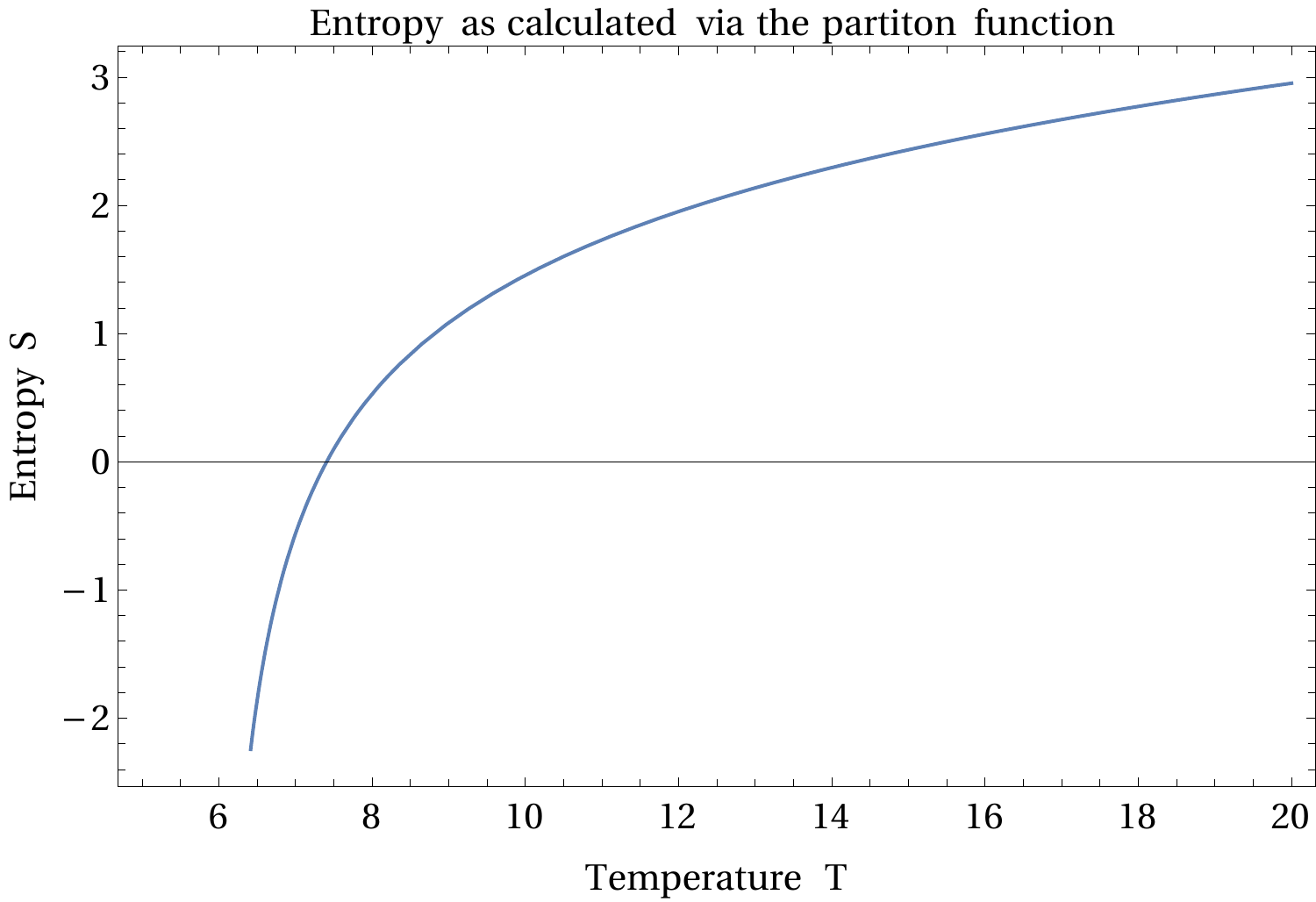}
 \end{subfigure}
    \caption{
    {\it Energy and entropy of the holographic model.}  Calculated via the partition function eq.~(\ref{partfunc}). \textit{Left:} The average energy. \textit{Right:} The entropy.\label{fig:Thermo}}
\end{figure}

In our model the mass of a five-dimensional, topological black hole is proportional to the winding number, $n$, 
\begin{equation}
    M(T)=\frac{3 \pi  L^2 n}{32 G_5}+\frac{3 \pi\,  m_t(T)\, n}{8 G_5}+\frac{3 \pi  n }{8 e^2 G_5}\log \left(\Lambda L\right) \, , %
\end{equation}
repeating eq.~\eqref{eq:Mass_equation} for convenience. 
The quantization of the mass implies the quantization of the entropy and the specific heat, just as in Bekenstein's black-hole-as-an-atom model.  These quantities can be obtained from a partition function, which can be calculated in analytic form
\be 
Z(T)\,=\, \frac{1}{1-e^{-\alpha(T)/T}}\, ,
\label{partfunc}
\ee
where
\be 
\alpha(T)\,=\, \frac{3 \pi  L^2 }{32 G_5}+\frac{3 \pi\,  m_t(T)\, }{8 G_5}+\frac{3 \pi   }{8 e^2 G_5}\log \left(\Lambda L\right)\, .
\ee
The average energy and entropy calculated from this partition function have the forms shown in figure~\ref{fig:Thermo}.

Qualitatively similar graphs can be obtained from the quantum-black-hole-as-an-atom model if the radius of the box $r$ in eq.\ref{Nmax} increases linearly with the temperature
\be 
r\,=\, \theta\,(T-T_0)\, ,
\ee
where $\theta$ is a constant, and $T_0$ is the value of the temperature for which $N_{max}\,=\ 0$. 

Although our model of a black hole is quite different from Bekenstein's atomic model, the thermodynamical quantities of the two models are very similar.  In both models the mass of the black hole is quantized, leading to quantization of the entropy and specific heat. Furthermore, there is qualitative agreement between the two models for the variation with respect to temperature of the average energy and the average entropy, providing that the containment box in Bekenstein's model varies linearly with temperature. In both models the black hole exists inside of a ``box''. In the atomic black hole model the black hole is placed in a box by fiat, a procedure which is necessary in order to render the partition function finite.  In our model the topological black hole exists in anti-de Sitter space, which due to the gravitational potential of such a space acts as a box of finite volume~\cite{Hawking:1975vcx,Gibbons:1976ue,York:1986yv}.
\paragraph{}
The major differences between our model and the atomic black hole model are the forms of the dependence of the masses on an integer and the forms of the partition functions. In our model the black hole mass depends linearly on an integer - the winding number.  In the atomic black hole model the black hole  horizon area depends linearly on an integer, which means that the black hole mass in that model is proportional to the square root of the integer. The canonical ensemble partition function in the atomic black hole model is a weighted finite sum over powers of even integers.  The sum which defines the partition function for this model must be finite in order to obtain a well-defined partition function.  The necessity of truncating the usual infinite series which defines a partition function is the reason the black hole in this model is placed in a box.  In our model the partition function can be expressed as an infinite series, which can be summed to obtain an analytic expression.   
\begin{figure}[htb]
\begin{center}
    \includegraphics[width=14cm]{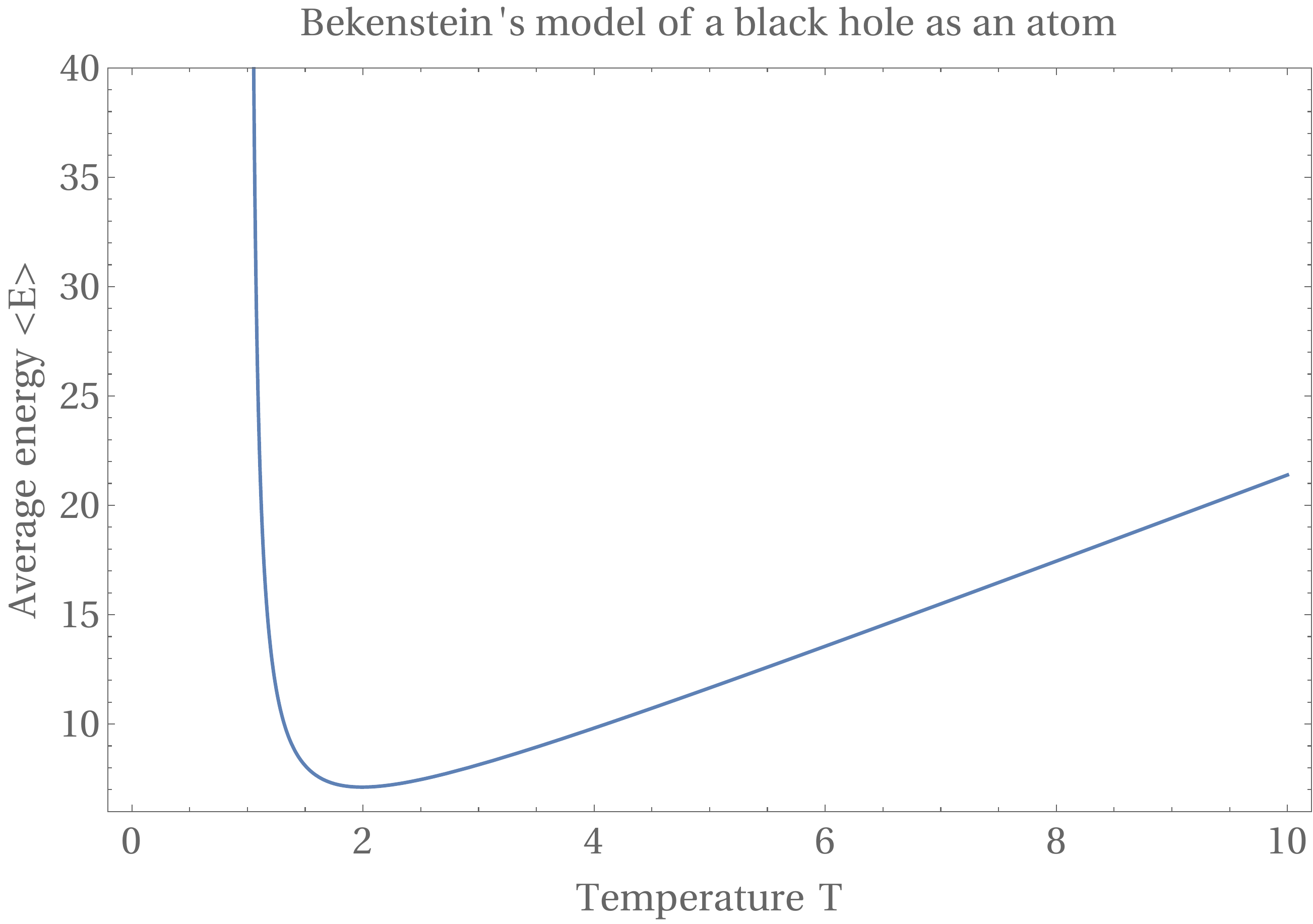}
    \end{center}
    \caption{
    \textit{Average energy:} The energy of a black hole in a box in Bekenstein's black-hole-as-an-atom model for a temperature-dependent box size. This figure is qualitatively similar to the average black hole energy obtained for our model (figure~\ref{fig:Thermo}). 
    \label{fig:Eatom}}
\end{figure}

\subsection{Kosterlitz - Thouless transitions}
In a system in a $2+1$-dimensional space with $O(2)$-symmetry the long-range order of the system is destroyed when the temperature of the system exceeds a critical value, resulting in the creation of defects (vortices).  For example, the free energy of a  global $O(2)$  scalar field  model
 is \cite{Stephens:2001sd}
 \be 
 F\,=\, (\pi\,\eta^2- 2\,T)\,\ln\left( \frac{L}{a}\right)\, ,
 \ee
 where $a\,\propto\,\frac{1}{\sqrt{\lambda\,\eta}}$, $\lambda$ is the coupling strength for a quartic interaction term and $\eta$ is the zero point for this term.
 In the two-dimensional X-Y model the free energy of a single vortex can be estimated from the relation
\be 
F\,=\,E - T\,S \, ,
\ee
where the energy of a vortex is
\be 
E\,=\, E_0+\pi n^2 J \ln\left(\frac{L}{a}\right)\, .
\ee
In this relation $E_0$ is the ground state energy of the system, which is the state when all of the rotors are completely aligned.
In this expression $n$ is a winding number, $J$ is the coupling strength between neighboring plane rotors on a two-dimensional square lattice, $L$ is the size of the system and $a$ is the lattice spacing. The entropy can be estimated by calculating the number of possible positions of a vortex on a square lattice with $L^2$ squares formed by nearest neighbors
\be 
S\,=\, k_B\,\ln\left( \frac{L^2}{a^2}\right)
\ee
The resulting free energy is
\be 
F\,=\, E_0 + \left(\pi\,J - 2\,k_B\,T\,\right)\ln\left(\frac{L}{a}\right)\, .
\ee
This heuristic argument shows that at a finite temperature a phase transition occurs in which vortex - anti-vortex pairs with energy
\be 
E_{v, \bar{v}}\, =\, 2\,E_c + \eta\, J\,\ln\left(\frac{L}{a}\right)
\label{KTfreeenrgy}
\ee
unbind, producing defects in the system.  This is the Kosterlitz - Thouless transition.  
\paragraph{}
In our model the Hawking - Page phase transition is described by eq.\ref{eq:DeltaFHawkingPage}, which contains a logarithmic term whose argument depends upon the size of the topological black hole. The free energy expression in the K - T transition, eq.\ref{KTfreeenrgy}, also contains a size-dependent logarithmic term.   Another parallel between the H - P transition in our model and the K-T transition is  that temperature at which the transitions occur depends upon the coupling strength ($r_h$ in our model is temperature dependent and coupling strength dependent).
\paragraph{}
In both models the energy of the primary constituent of the system, a single vortex in the  X - Y model, a black hole in our model, depend on a winding number.  However, in the X - Y model the energy is proportional to $n^2$, but in our model the energy (mass) of the black hole is linearly proportional $n$. Another difference between the K - T transition and the transition described in our model is that the number of H - P transitions which can occur depends on the coupling strength. As shown in Section 4, the number of H - P phase transitions can be 0, 1, or 2.  For the latter case, it is possible to form either a large black hole (large horizon radius) or a small black hole at a given temperature.

\subsection{Hagedorn transitions}
The Hagedorn model of strong interactions at high energies is a statistical - thermodynamical model which treats resonances of strongly interacting particles as ``fireballs'', which are composed of smaller ``fireballs'', which are composed of still smaller ``fireballs'', and so on \cite{Hagedorn:1965st}.  This approach results in an asymptotic bootstrap equation for the density of states.  The solution of this equation predicts that the mass spectrum grows exponentially
\be 
\rho(m)\, \stackrel{m\to\infty}{\longrightarrow}\, C\,m^{-5/2}\,\exp\left(\frac{m}{T_H}\right)
\label{massdens}
\ee
where $C$ is a constant and $T_H$ is the highest possible temperature for strong interactions.  Above this temperature  the partition function diverges.  The physical interpretation of this effect is that at $T_H$ a phase transition occurs in which quarks become deconfined.  This is the Hagedorn transition.  The expression for the density of states was generalized in \cite{Frautschi:1971ij}
\be 
\rho(m)\, \stackrel{m\to\infty}{\longrightarrow}\, C\,f(m)\,\exp\left(\frac{m^p}{T_H}\right)
\label{genmassdns}
\ee
where $f(m)$ is a polynomial in $m$.  A gas of superstrings has  a density of states with same general form as the one in eq.\ref{genmassdns}  \cite{Harms:1992nb} with $p\,=\,2$. The Hagedorn transition in $SU(N)$ gauge theory for $N\,\to\,\infty$ is connected to the  string theory description of  black hole formation in anti-de Sitter $(AdS)$ space \cite{Aharony:1999ti}.  The deconfinement phase transition of a supersymmetric gauge theory on the $AdS$ boundary is dual to the Hawking - Page transition during which a black hole is formed in the bulk \cite{Witten:1998zw}.  The topological black hole in our model is dual to $\mathcal{N}=4$ Super-Yang-Mills theory~(SYM) in Minkowski space coupled to an external $SU(2)$ gauge field $F$ associated with an $SU(2)$ subgroup of the $SU(4)$ R-symmetry of the $\mathcal{N}=4$ SYM theory, see e.g.~\cite{Son:2006em,Behrndt:1998jd,Cvetic:1999ne,Gubser:1998jb,Chamblin:1999tk}.  The dual state is a strongly coupled quantum many-body system at non-zero temperature.   Thus the dual to our topological black hole is an example of a system which can undergo a Hagedorn transition.  However, in our model two Hawking - Page transitions are possible, depending on the strength of the Skyrme coupling constant.   This means that there are two states of deconfinement in the dual conformal field theory, a lower energy state and a higher energy state.  In quantum chromodynamics the deconfined state is usually considered to be a `liquid' composed of free or nearly free quarks.  The interpretation of a second state of deconfinement has not yet been worked out.


%

\section{Discussion}
\label{sec:discussion}
The topological black hole in our model possesses several interesting features.  The mass of the black hole and therefore its energy and its entropy  are quantized.  The quantization integer is the winding number, which parameterizes the solutions of the Einstein-Skyrme field equations.  The Skyrme field possessed by the black hole determines the number of Hawking - Page (H - P) transitions which the system can undergo.  The number of transitions can be zero, one or two, depending on the strength of the Skyrme field coupling constant and on the winding number.  The fact that two H - P phase transitions at different temperatures can occur presents an interesting challenge to the holographic interpretation of these transitions.  An H - P transition in the bulk is usually interpreted as the (de-)confinement transition of a dual, strongly-interacting system on the $AdS$ boundary.  
Now the Skyrmion (holographically dual to the $SU(2)$ gauge sector on the field theory side) leads to the existence of a second transition at particular values of the Skyrme coupling $e$ and winding numbers $n$ of the $SU(2)$ field configurations.
This raises the obvious question: 
What type of (topological) physical principle in the dual field theory leads to the second phase transition? 
It will be interesting to translate the observed phenomenology of our model into a more concise condensed matter context.

\section*{Acknowledgements}
This research was supported in part by DOE grant DE-SC-0012447.

\bibliographystyle{JHEP}
\bibliography{geo_pub}

\providecommand{\href}[2]{#2}\begingroup\raggedright\begin{thebibliography}{10}

\bibitem{Cartwright:2020yoc}
C.~Cartwright, B.~Harms and M.~Kaminski, \emph{{ Topological or Rotational
  Non-Abelian Gauge Fields from Einstein-Skyrme Holography}},
  \href{https://doi.org/10.1007/JHEP03(2021)229}{\emph{JHEP} {\bfseries 03}
  (2021) 229} [\href{https://arxiv.org/abs/2010.03578}{{\ttfamily
  2010.03578}}].

\bibitem{Maldacena:1997re}
J.~M. Maldacena, \emph{{The Large N limit of superconformal field theories and
  supergravity}}, \href{https://doi.org/10.1023/A:1026654312961,
  10.4310/ATMP.1998.v2.n2.a1}{\emph{Int. J. Theor. Phys.} {\bfseries 38} (1999)
  1113} [\href{https://arxiv.org/abs/hep-th/9711200}{{\ttfamily
  hep-th/9711200}}].

\bibitem{Witten:1998zw}
E.~Witten, \emph{{Anti-de Sitter space, thermal phase transition, and
  confinement in gauge theories}},
  \href{https://doi.org/10.4310/ATMP.1998.v2.n3.a3}{\emph{Adv. Theor. Math.
  Phys.} {\bfseries 2} (1998) 505}
  [\href{https://arxiv.org/abs/hep-th/9803131}{{\ttfamily hep-th/9803131}}].

\bibitem{haldane2016ground}
F.~D.~M. Haldane, \emph{Ground state properties of antiferromagnetic chains
  with unrestricted spin: Integer spin chains as realisations of the o(3)
  non-linear sigma model},  2016.

\bibitem{PhysRevB.65.165113}
X.-G. Wen, \emph{Quantum orders and symmetric spin liquids},
  \href{https://doi.org/10.1103/PhysRevB.65.165113}{\emph{Phys. Rev. B}
  {\bfseries 65} (2002) 165113}.

\bibitem{doi:10.1126/science.1091806}
T.~Senthil, A.~Vishwanath, L.~Balents, S.~Sachdev and M.~P.~A. Fisher,
  \emph{Deconfined quantum critical points},
  \href{https://doi.org/10.1126/science.1091806}{\emph{Science} {\bfseries 303}
  (2004) 1490}
  [\href{https://arxiv.org/abs/https://www.science.org/doi/pdf/10.1126/science.1091806}{{\ttfamily
  https://www.science.org/doi/pdf/10.1126/science.1091806}}].

\bibitem{Kharzeev:2004ey}
D.~Kharzeev, \emph{Parity violation in hot qcd: Why it can happen, and how to
  look for it},
  \href{https://doi.org/10.1016/j.physletb.2005.11.075}{\emph{Phys. Lett. B}
  {\bfseries 633} (2006) 260}
  [\href{https://arxiv.org/abs/hep-ph/0406125}{{\ttfamily hep-ph/0406125}}].

\bibitem{Ghosh:2021naw}
J.~K. Ghosh, S.~Grieninger, K.~Landsteiner and S.~Morales-Tejera, \emph{{Is the
  chiral magnetic effect fast enough?}},
  \href{https://doi.org/10.1103/PhysRevD.104.046009}{\emph{Phys. Rev. D}
  {\bfseries 104} (2021) 046009}
  [\href{https://arxiv.org/abs/2105.05855}{{\ttfamily 2105.05855}}].

\bibitem{Cartwright:2021maz}
C.~Cartwright, M.~Kaminski and B.~Schenke, \emph{{Energy dependence of the
  chiral magnetic effect in expanding holographic plasma}},
  \href{https://arxiv.org/abs/2112.13857}{{\ttfamily 2112.13857}}.

\bibitem{Cartwright:2020qov}
C.~Cartwright, \emph{Entropy production far from equilibrium in a chiral
  charged plasma in the presence of external electromagnetic fields},
  \href{https://arxiv.org/abs/2003.04325}{{\ttfamily 2003.04325}}.

\bibitem{Taylor:2000xw}
M.~Taylor, \emph{{More on counterterms in the gravitational action and
  anomalies}},  \href{https://arxiv.org/abs/hep-th/0002125}{{\ttfamily
  hep-th/0002125}}.

\bibitem{Gross:1982cv}
D.~J. Gross, M.~J. Perry and L.~G. Yaffe, \emph{Instability of flat space at
  finite temperature},
  \href{https://doi.org/10.1103/PhysRevD.25.330}{\emph{Phys. Rev. D} {\bfseries
  25} (1982) 330}.

\bibitem{Hawking:1975vcx}
S.~W. Hawking, \emph{Particle creation by black holes},
  \href{https://doi.org/10.1007/BF02345020}{\emph{Commun. Math. Phys.}
  {\bfseries 43} (1975) 199}.

\bibitem{Gibbons:1976ue}
G.~W. Gibbons and S.~W. Hawking, \emph{Action integrals and partition functions
  in quantum gravity},
  \href{https://doi.org/10.1103/PhysRevD.15.2752}{\emph{Phys. Rev. D}
  {\bfseries 15} (1977) 2752}.

\bibitem{York:1986yv}
J.~W. York, \emph{Black-hole thermodynamics and the euclidean einstein action},
  \href{https://doi.org/10.1103/PhysRevD.33.2092}{\emph{Phys. Rev. D}
  {\bfseries 33} (1986) 2092}.

\bibitem{Chamblin:1999hg}
A.~Chamblin, R.~Emparan, C.~V. Johnson and R.~C. Myers, \emph{{Holography,
  thermodynamics and fluctuations of charged AdS black holes}},
  \href{https://doi.org/10.1103/PhysRevD.60.104026}{\emph{Phys. Rev. D}
  {\bfseries 60} (1999) 104026}
  [\href{https://arxiv.org/abs/hep-th/9904197}{{\ttfamily hep-th/9904197}}].

\bibitem{Balasubramanian:1999re}
V.~Balasubramanian and P.~Kraus, \emph{A stress tensor for anti-de sitter
  gravity}, \href{https://doi.org/10.1007/s002200050764}{\emph{Commun. Math.
  Phys.} {\bfseries 208} (1999) 413}
  [\href{https://arxiv.org/abs/hep-th/9902121}{{\ttfamily hep-th/9902121}}].

\bibitem{hawking1982}
S.~W. Hawking and D.~N. Page, \emph{Thermodynamics of black holes in anti-de
  sitter space}, {\emph{Comm. Math. Phys.} {\bfseries 87} (1982) 577}.

\bibitem{Canfora:2013osa}
F.~Canfora and H.~Maeda, \emph{{Hedgehog ansatz and its generalization for
  self-gravitating Skyrmions}},
  \href{https://doi.org/10.1103/PhysRevD.87.084049}{\emph{Phys. Rev.}
  {\bfseries D87} (2013) 084049}
  [\href{https://arxiv.org/abs/1302.3232}{{\ttfamily 1302.3232}}].

\bibitem{Ayon-Beato:2015eca}
E.~Ayon-Beato, F.~Canfora and J.~Zanelli, \emph{{Analytic self-gravitating
  Skyrmions, cosmological bounces and AdS wormholes}},
  \href{https://doi.org/10.1016/j.physletb.2015.11.065}{\emph{Phys. Lett. B}
  {\bfseries 752} (2016) 201}
  [\href{https://arxiv.org/abs/1509.02659}{{\ttfamily 1509.02659}}].

\bibitem{Canfora:2017yio}
F.~Canfora, S.~H. Oh and P.~Salgado-Rebolledo, \emph{Gravitational catalysis of
  merons in einstein-yang-mills theory},
  \href{https://doi.org/10.1103/PhysRevD.96.084038}{\emph{Phys. Rev. D}
  {\bfseries 96} (2017) 084038}
  [\href{https://arxiv.org/abs/1710.00133}{{\ttfamily 1710.00133}}].

\bibitem{Canfora:2018ppu}
F.~Canfora, A.~Gomberoff, S.~H. Oh, F.~Rojas and P.~Salgado-Rebolledo,
  \emph{Meronic einstein-yang-mills black hole in 5d and gravitational spin
  from isospin effect},
  \href{https://doi.org/10.1007/JHEP06(2019)081}{\emph{JHEP} {\bfseries 06}
  (2019) 081} [\href{https://arxiv.org/abs/1812.11231}{{\ttfamily
  1812.11231}}].

\bibitem{Ayon-Beato:2019tvu}
E.~Ay\'on-Beato, F.~Canfora, M.~Lagos, J.~Oliva and A.~Vera, \emph{Analytic
  self-gravitating $4$-baryons, traversable nut-ads wormholes, flat space-time
  multi-skyrmions at finite volume and a novel transition in the $su(3)$-skyrme
  model}, \href{https://doi.org/10.1140/epjc/s10052-020-7926-6}{\emph{Eur.
  Phys. J. C} {\bfseries 80} (2020) 384}
  [\href{https://arxiv.org/abs/1909.00540}{{\ttfamily 1909.00540}}].

\bibitem{Ipinza:2020xgc}
M.~Ipinza and P.~Salgado-Rebolledo, \emph{{Meron-like topological solitons in
  massive Yang-Mills theory and the Skyrme model}},
  \href{https://arxiv.org/abs/2005.04920}{{\ttfamily 2005.04920}}.

\bibitem{RevModPhys.51.461}
A.~Actor, \emph{Classical solutions of $\mathrm{SU}(2)$ yang---mills theories},
  \href{https://doi.org/10.1103/RevModPhys.51.461}{\emph{Rev. Mod. Phys.}
  {\bfseries 51} (1979) 461}.

\bibitem{Kunimasa:1967GT}
T.~Kunimasa and T.~Gotō, \emph{{Generalization of the Stueckelberg Formalism
  to the Massive Yang-Mills Field}},
  \href{https://doi.org/10.1143/PTP.37.452}{\emph{Progress of Theoretical
  Physics} {\bfseries 37} (1967) 452}.

\bibitem{Shizuya:1975ek}
K.-i. Shizuya, \emph{{Quantization of the Massive Yang-Mills Field in Arbitrary
  Gauges}}, \href{https://doi.org/10.1016/0550-3213(75)90492-7}{\emph{Nucl.
  Phys. B} {\bfseries 94} (1975) 260}.

\bibitem{Klebanov:2002gr}
I.~R. Klebanov, P.~Ouyang and E.~Witten, \emph{{A Gravity dual of the chiral
  anomaly}}, \href{https://doi.org/10.1103/PhysRevD.65.105007}{\emph{Phys. Rev.
  D} {\bfseries 65} (2002) 105007}
  [\href{https://arxiv.org/abs/hep-th/0202056}{{\ttfamily hep-th/0202056}}].

\bibitem{Son:2006em}
D.~T. Son and A.~O. Starinets, \emph{{Hydrodynamics of r-charged black holes}},
  \href{https://doi.org/10.1088/1126-6708/2006/03/052}{\emph{JHEP} {\bfseries
  03} (2006) 052} [\href{https://arxiv.org/abs/hep-th/0601157}{{\ttfamily
  hep-th/0601157}}].

\bibitem{Behrndt:1998jd}
K.~Behrndt, M.~Cvetic and W.~Sabra, \emph{{Nonextreme black holes of
  five-dimensional N=2 AdS supergravity}},
  \href{https://doi.org/10.1016/S0550-3213(99)00243-6}{\emph{Nucl. Phys. B}
  {\bfseries 553} (1999) 317}
  [\href{https://arxiv.org/abs/hep-th/9810227}{{\ttfamily hep-th/9810227}}].

\bibitem{Cvetic:1999ne}
M.~Cvetic and S.~S. Gubser, \emph{Phases of r charged black holes, spinning
  branes and strongly coupled gauge theories},
  \href{https://doi.org/10.1088/1126-6708/1999/04/024}{\emph{JHEP} {\bfseries
  04} (1999) 024} [\href{https://arxiv.org/abs/hep-th/9902195}{{\ttfamily
  hep-th/9902195}}].

\bibitem{Gubser:1998jb}
S.~S. Gubser, \emph{{Thermodynamics of spinning D3-branes}},
  \href{https://doi.org/10.1016/S0550-3213(99)00194-7}{\emph{Nucl. Phys. B}
  {\bfseries 551} (1999) 667}
  [\href{https://arxiv.org/abs/hep-th/9810225}{{\ttfamily hep-th/9810225}}].

\bibitem{Chamblin:1999tk}
A.~Chamblin, R.~Emparan, C.~V. Johnson and R.~C. Myers, \emph{{Charged AdS
  black holes and catastrophic holography}},
  \href{https://doi.org/10.1103/PhysRevD.60.064018}{\emph{Phys. Rev. D}
  {\bfseries 60} (1999) 064018}
  [\href{https://arxiv.org/abs/hep-th/9902170}{{\ttfamily hep-th/9902170}}].

\bibitem{Bera:2017}
A.~K. Bera, B.~Lake, F.~H.~L. Essler, L.~Vanderstraeten, C.~Hubig,
  U.~Schollw\"ock et~al., \emph{Spinon confinement in a quasi-one-dimensional
  anisotropic heisenberg magnet},
  \href{https://doi.org/10.1103/PhysRevB.96.054423}{\emph{Phys. Rev. B}
  {\bfseries 96} (2017) 054423}.

\bibitem{Lake:2009}
B.~Lake, A.~M. Tsvelik, S.~Notbohm, D.~Alan~Tennant, T.~G. Perring, M.~Reehuis
  et~al., \emph{Confinement of fractional quantum number particles in a
  condensed-matter system},
  \href{https://doi.org/10.1038/nphys1462}{\emph{Nature Physics} {\bfseries 6}
  (2009) 50–55}.

\bibitem{Vestergren:2005in}
A.~Vestergren and J.~Lidmar, \emph{Topological order and the deconfinement
  transition in the (2+1) dimensional compact abelian higgs model},
  \href{https://doi.org/10.1103/PhysRevB.72.174515}{\emph{Phys. Rev. B}
  {\bfseries 72} (2005) 174515}
  [\href{https://arxiv.org/abs/cond-mat/0502533}{{\ttfamily
  cond-mat/0502533}}].

\bibitem{Bekenstein:1995ju}
J.~D. Bekenstein and V.~F. Mukhanov, \emph{{Spectroscopy of the quantum black
  hole}}, \href{https://doi.org/10.1016/0370-2693(95)01148-J}{\emph{Phys. Lett.
  B} {\bfseries 360} (1995) 7}
  [\href{https://arxiv.org/abs/gr-qc/9505012}{{\ttfamily gr-qc/9505012}}].

\bibitem{Bekenstein:1997bt}
J.~D. Bekenstein, \emph{{Quantum black holes as atoms}},  in \emph{8th Marcel
  Grossmann Meeting on Recent Developments in Theoretical and Experimental
  General Relativity, Gravitation and Relativistic Field Theories (MG 8)},
  pp.~92--111, 6, 1997, \href{https://arxiv.org/abs/gr-qc/9710076}{{\ttfamily
  gr-qc/9710076}}.

\bibitem{Stephens:2001sd}
G.~J. Stephens and B.~L. Hu, \emph{{Notes on black hole phase transitions}},
  \href{https://doi.org/10.1023/A:1012930019453}{\emph{Int. J. Theor. Phys.}
  {\bfseries 40} (2001) 2183}
  [\href{https://arxiv.org/abs/gr-qc/0102052}{{\ttfamily gr-qc/0102052}}].

\bibitem{Hagedorn:1965st}
R.~Hagedorn, \emph{{Statistical thermodynamics of strong interactions at
  high-energies}}, {\emph{Nuovo Cim. Suppl.} {\bfseries 3} (1965) 147}.

\bibitem{Frautschi:1971ij}
S.~C. Frautschi, \emph{{Statistical bootstrap model of hadrons}},
  \href{https://doi.org/10.1103/PhysRevD.3.2821}{\emph{Phys. Rev. D} {\bfseries
  3} (1971) 2821}.

\bibitem{Harms:1992nb}
B.~Harms and Y.~Leblanc, \emph{{Statistical mechanics of black holes}},
  \href{https://doi.org/10.1103/PhysRevD.46.2334}{\emph{Phys. Rev. D}
  {\bfseries 46} (1992) 2334}
  [\href{https://arxiv.org/abs/hep-th/9205021}{{\ttfamily hep-th/9205021}}].

\bibitem{Aharony:1999ti}
O.~Aharony, S.~S. Gubser, J.~M. Maldacena, H.~Ooguri and Y.~Oz, \emph{{Large N
  field theories, string theory and gravity}},
  \href{https://doi.org/10.1016/S0370-1573(99)00083-6}{\emph{Phys. Rept.}
  {\bfseries 323} (2000) 183}
  [\href{https://arxiv.org/abs/hep-th/9905111}{{\ttfamily hep-th/9905111}}].

\end{thebibliography}\endgroup



\providecommand{\href}[2]{#2}\begingroup\raggedright\endgroup



\providecommand{\href}[2]{#2}\begingroup\raggedright\endgroup



\providecommand{\href}[2]{#2}\begingroup\raggedright\endgroup
\end{document}